\documentclass{article}
\usepackage{iclr2027_conference,times}
\iclrfinalcopy

\usepackage{amsmath,amsfonts,bm}

\def\eqref#1{equation~\ref{#1}}

\def\1{\bm{1}}

\DeclareMathAlphabet{\mathsfit}{\encodingdefault}{\sfdefault}{m}{sl}
\SetMathAlphabet{\mathsfit}{bold}{\encodingdefault}{\sfdefault}{bx}{n}

\usepackage[utf8]{inputenc}
\usepackage[T1]{fontenc}
\usepackage{hyperref}
\usepackage{url}
\usepackage{booktabs}
\usepackage{graphicx}
\usepackage{float}
\usepackage{placeins}
\usepackage{amsmath}
\usepackage{amsfonts}
\usepackage{nicefrac}
\usepackage{microtype}
\usepackage{xcolor}
\usepackage{fvextra}
\usepackage{fontawesome5}
\newcommand{\promptfont}{\fontencoding{T1}\fontfamily{cmtt}\fontseries{m}\fontshape{n}\fontsize{8}{10}\selectfont}
\usepackage{enumitem}
\usepackage{tcolorbox}
\tcbuselibrary{breakable,skins}
\definecolor{PromptBackground}{HTML}{F7F8FA}
\definecolor{PromptBorder}{HTML}{D4D8DE}
\definecolor{PromptInk}{HTML}{303B49}
\newtcolorbox{SystemPromptBox}[1]{enhanced,breakable,
  colback=PromptBackground,colframe=PromptBorder,colbacktitle=PromptBackground,
  coltitle=PromptInk,fonttitle=\sffamily\bfseries\small,
  title={#1},title after break={#1\enspace\normalfont\small(continued)},
  boxrule=0.45pt,arc=1.5mm,left=10pt,right=10pt,top=8pt,bottom=8pt,
  toptitle=7pt,bottomtitle=7pt,pad at break*=3mm,
  before upper={\small\setlength{\parindent}{0pt}\setlength{\parskip}{2pt}%
    \setlist[itemize]{leftmargin=1.35em,itemsep=1pt,parsep=1pt,topsep=2pt}%
    \setlist[enumerate]{leftmargin=1.6em,itemsep=1pt,parsep=1pt,topsep=2pt}%
    \emergencystretch=2em}}
\makeatletter
\newcommand{\PromptHeading}{\@startsection{paragraph}{4}{0pt}%
  {7pt}{3pt}{\color{PromptInk}\sffamily\bfseries}*}
\newcommand{\PromptSubheading}{\@startsection{paragraph}{4}{0pt}%
  {5pt}{2pt}{\bfseries}*}
\makeatother
\newcommand{\PromptSourceTitle}[1]{{\footnotesize\itshape #1}\par}
\DefineVerbatimEnvironment{PromptCode}{Verbatim}{fontsize=\footnotesize,
  formatcom=\promptfont,breaklines=true,breakanywhere=true,breaksymbolleft={},
  xleftmargin=4pt,xrightmargin=4pt,frame=leftline,rulecolor=\color{PromptBorder}}

\definecolor{ProjectBlue}{HTML}{1A73E8}
\definecolor{GitHubBlack}{HTML}{24292F}
\definecolor{HFOrange}{HTML}{FF9D00}
\title{VISTA: An End-to-End Benchmark for Visual Spec-to-App Coding Agents}

\author{%
\hbox to\textwidth{\hfil\normalfont
\begin{tabular}{@{}p{0.44\textwidth}@{\hspace{0.04\textwidth}}p{0.44\textwidth}@{}}
\textbf{JunJia Guo} & \textbf{Yuhang Yao} \\
University of Arizona & Zoom \\
\texttt{junjiaguo@arizona.edu} & \texttt{yuhangyao8@gmail.com} \\[1em]
\textbf{Chang Li} & \textbf{Zimeng Pan} \\
George Washington University & Google \\
\texttt{cl885@gwmail.gwu.edu} & \texttt{justinzmpan@gmail.com} \\[1em]
\textbf{Jiawei (Joe) Zhou} & \textbf{Jingdi Chen} \\
Stony Brook University & University of Arizona \\
\texttt{jiawei.zhou.1@stonybrook.edu} & \texttt{jingdic@arizona.edu}
\end{tabular}
\hfil}%
}

\begin{document}

\maketitle
\thispagestyle{empty}
\lhead{}

\begin{center} \vspace{-0.5em} \href{https://kaboider.github.io/VIS_APP/} {\textcolor{ProjectBlue}{\faIcon{globe}}~Project Page} \quad \href{https://github.com/kaboider/VISTA_Bench} {\textcolor{GitHubBlack}{\faIcon{github}}~Code} \quad \href{https://huggingface.co/datasets/JunJiaGuo/VIS-APP-Bench} {\textcolor{HFOrange}{\faIcon{database}}~Dataset} \vspace{0.5em} \end{center}

\begin{abstract}
Coding agents can now build applications from design mockups, but a screen that
looks right is not an application that works. Existing benchmarks either score
visual reconstruction without a backend or test full-stack functionality from
textual requirements, and few tie scores to the individual components a design
requires. We introduce VISTA (VIsual Spec-To-App), an end-to-end benchmark in
which coding agents turn multi-page design handoffs (Figma renders, structure,
and textual requirements) into runnable full-stack Web and Mobile
(Android) applications. VISTA provides 8{,}487 human-annotated interactive
components across 18 applications (126 Web pages and 576 Android screens) and an
executable evaluator that locates annotated components in the running
application and probes their interactions, yielding localization, behavior, and
joint scores traceable to individual design requirements. Across 14 deployed
coding-agent systems, the best joint score is 0.553 on Web and 0.378 on Mobile,
and 13 of 14 systems score lower on behavior than on localization on Web.
Component probes detect unresponsive controls in 90.1\% of audited Web
deliveries. Development traces show where the process departs from the task:
44.5\% of trajectories never open the design screenshots, 73.0\% omit the
prescribed self-audit, and 8.3\% report success after a failed check. VISTA
makes the gap between rendering a design and delivering a working application
measurable at the level of individual components. Code is available at
\href{https://github.com/kaboider/VISTA_Bench}.
\end{abstract}

\section{Introduction}

Building an application from a design requires translating page layouts and components into connected interfaces, backend logic, and persistent data. We study this workflow as \emph{full-stack design-to-app}: constructing and launching a web or mobile application from Figma designs and accompanying product requirements. A task-creation form, for example, must preserve the specified interface while connecting user input to a stored record. These requirements fail independently: a screen can match the design yet lack working submission logic, and a functional implementation can depart from the design. Evaluation therefore requires checking, at each required component, both where it appears and whether it works.

Existing benchmarks cover parts of this workflow. Visual code-generation benchmarks score reconstruction and front-end interaction from screenshots or prototypes, without requiring a backend or database \citep{si2024design2code,xiao2026designbench,xiao2026interaction2code}. Full-stack benchmarks test frontend, backend, and database functionality, but start from textual requirements and do not assess fidelity to a supplied design \citep{lu2025webgenbench,lu2026fullstackagent}. Benchmarks that combine visual input with runtime checks evaluate interactions inferred from the interface on Web alone \citep{chen2026ui2app,zhao2026longwebbench}. What is missing is an explicit reference that connects each component of a design to an executable check on the delivered application, for both Web and mobile platforms.

\begin{figure}[t]
\centering
\includegraphics[width=\linewidth]{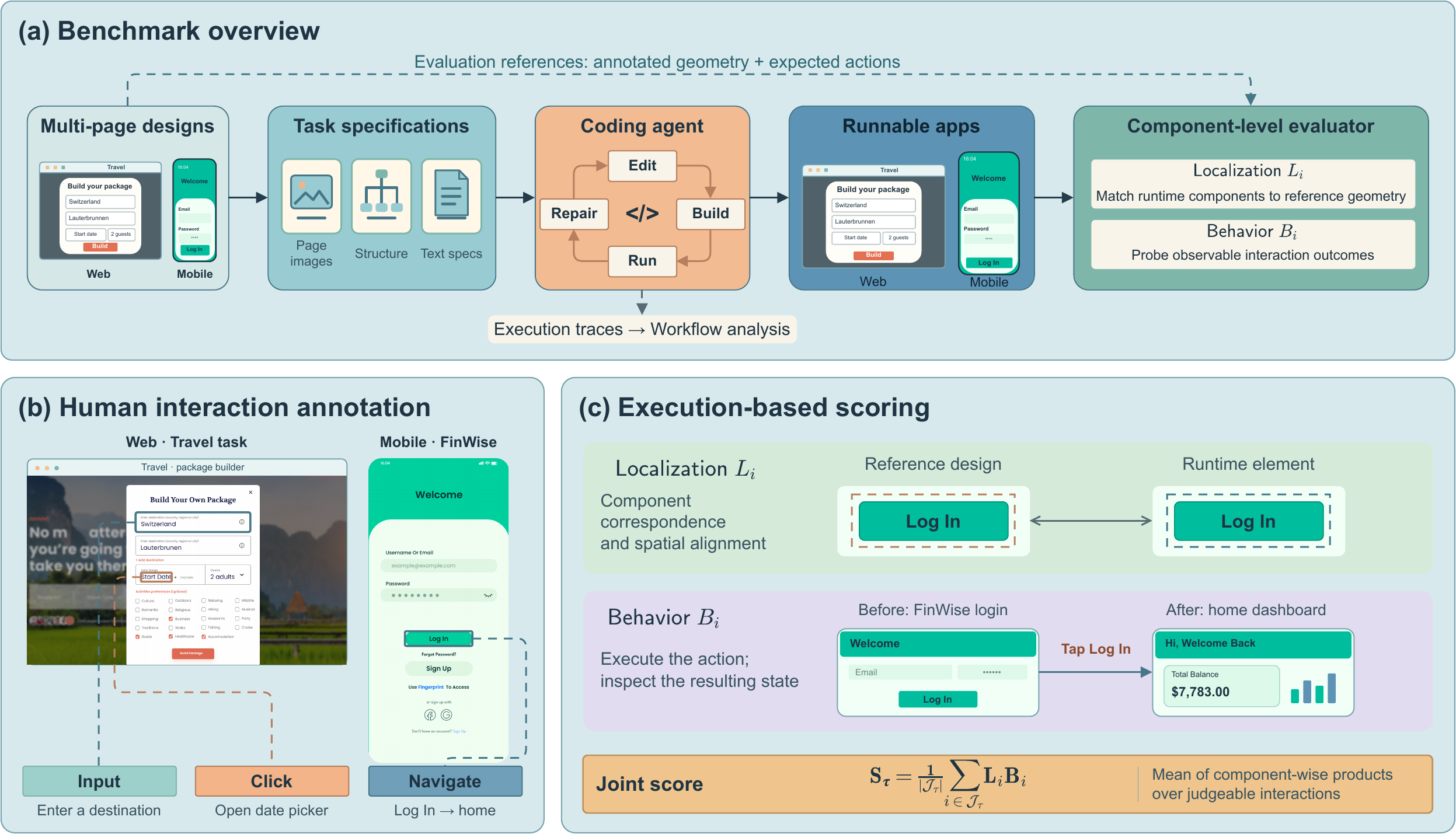}
\caption{Overview of the VISTA benchmark pipeline, from design collection and
annotation to agent execution, application evaluation, and workflow analysis.}
\label{fig:construction-draft}
\end{figure}

We introduce \textbf{VISTA}, a benchmark for Figma-to-full-stack application development on Web and Mobile (Figure~\ref{fig:construction-draft}). It contains 18 applications spanning 126 Web pages and 576 Android screens, with 8{,}487 manually annotated interactive components. Each annotation specifies where a control belongs, which interaction it must support, and, where applicable, its navigation target. Agents build and launch multi-page applications with frontend, backend, and database components in a self-contained environment. The evaluator matches annotated controls to runtime elements and executes a probe selected by each control's interaction type, producing localization and behavior scores that trace every unit of credit to a design requirement and its execution evidence. Development trajectories are retained alongside the scores.

Evaluating 14 deployed coding-agent systems shows that current agents render designs more reliably than they implement them (Table~\ref{tab:evaluation-web-mobile}). On Web, 13 of 14 systems score lower on behavior than on localization (mean $B=0.543$ versus $L=0.642$). The best joint scores are 0.553 on Web (Grok 4.6) and 0.378 on Mobile (GPT-6-astra), leaving substantial headroom on both tracks, and platform rankings diverge: GPT-6-astra ranks seventh on Web but first on Mobile. Component probes detect unresponsive controls in 90.1\% of audited Web deliveries, and Mobile failures extend to database initialization and launch (Section~\ref{sec:failure-examples}). Development traces expose process-level gaps (Section~\ref{sec:alignment}): 44.5\% of trajectories never open the design screenshots, 73.0\% omit the prescribed self-audit, and 8.3\% report success after a failed check.

Our contributions are:
\begin{itemize}
\item \textbf{A Figma-to-full-stack benchmark for Web and Mobile.} VISTA turns design handoffs into application-construction tasks that require integrated frontend, backend, and database implementations across 18 applications.
\item \textbf{Component-level ground truth with executable evaluation.} 8{,}487 human-annotated components define reference locations, interaction types, and navigation targets; the evaluator checks each against the running application, yielding localization and behavior scores traceable to individual requirements and validated against blind human judgments.
\item \textbf{Evidence that agents render designs better than they implement them.} Behavior trails localization for 13 of 14 systems, rankings diverge across platforms, and component probes expose unresponsive controls and deployment failures.
\item \textbf{A trajectory-level diagnosis.} Agents frequently bypass the design evidence and self-audit they are given, and report success beyond what their own checks show; richer design inputs increase agent activity without measurable score gains.
\end{itemize}

\section{Related Work}
\label{sec:related}

\subsection{Benchmarks for web and application development}

Visual code-generation benchmarks assess reconstruction from interface references.
Design2Code evaluates screenshot reconstruction, while DesignBench extends evaluation
to generation, editing, and repair across front-end frameworks
\citep{si2024design2code,xiao2026designbench}. Interaction2Code adds before-and-after
interaction prototypes; MRWeb supplies structured resource lists for links, images,
and backend routes \citep{xiao2026interaction2code,wan2024mrweb}. UI2App and
LongWebBench further evaluate inferred interactions and cross-route state, and
long-page structure with executable user goals, respectively
\citep{chen2026ui2app,zhao2026longwebbench}.

Benchmarks starting from textual requirements emphasize application functionality.
WebGen-Bench combines browser-executed tests with appearance assessment, and
MiniAppBench evaluates intent, presentation, and dynamic behavior
\citep{lu2025webgenbench,zhang2026miniappbench}. FullStack-Bench tests frontend,
backend, and database functionality, including database interactions accompanying
user actions \citep{lu2026fullstackagent}. Table~\ref{tab:benchmarks} compares the
platform coverage, inputs, and evaluation settings. None of these benchmarks
combines design-grounded, component-level ground truth with full-stack execution,
and all target Web only. VISTA covers both browser-based Web applications and
installable Android apps, with a shared component-level score formulation and
platform-specific execution checks. Its human-annotated component requirements
from Figma design handoffs specify what to inspect and which interactions to
execute, connecting each score to explicit ground truth and runtime evidence.

\begin{table}[t]
\centering
\small
\setlength{\tabcolsep}{6pt}
\caption{Comparison of application-generation benchmarks by target platform, supplied inputs, and interaction evaluation. V: supplied visual references; S: supplied structured design/resource information; I: runtime interaction evaluation. $\surd$ indicates inclusion; -- indicates outside the described protocol.}
\label{tab:benchmarks}
\begin{tabular}{@{}lccccc@{}}
\toprule
Benchmark & Web & Mobile & V & S & I \\
\midrule
Design2Code \citep{si2024design2code} & $\surd$ & -- & $\surd$ & -- & -- \\
DesignBench \citep{xiao2026designbench} & $\surd$ & -- & $\surd$ & -- & -- \\
Interaction2Code \citep{xiao2026interaction2code} & $\surd$ & -- & $\surd$ & -- & $\surd$ \\
MRWeb \citep{wan2024mrweb} & $\surd$ & -- & $\surd$ & $\surd$ & -- \\
WebGen-Bench \citep{lu2025webgenbench} & $\surd$ & -- & -- & -- & $\surd$ \\
MiniAppBench \citep{zhang2026miniappbench} & $\surd$ & -- & -- & -- & $\surd$ \\
FullStack-Bench \citep{lu2026fullstackagent} & $\surd$ & -- & -- & -- & $\surd$ \\
UI2App \citep{chen2026ui2app} & $\surd$ & -- & $\surd$ & -- & $\surd$ \\
LongWebBench \citep{zhao2026longwebbench} & $\surd$ & -- & $\surd$ & -- & $\surd$ \\
\textbf{VISTA} & $\surd$ & $\surd$ & $\surd$ & $\surd$ & $\surd$ \\
\bottomrule
\end{tabular}
\par\smallskip
\begin{minipage}{0.95\linewidth}
\footnotesize
Web and Mobile denote application runtimes; VISTA's Mobile track produces installable Android apps.
S denotes design/resource information supplied during generation, excluding evaluator-only DOM access and source code supplied for repair. Interaction2Code uses v3; DesignBench uses v1.
\end{minipage}
\end{table}

\subsection{Coding agents and process evaluation}

Coding-agent research examines both task completion and the interfaces through
which agents act. SWE-bench evaluates repository issue resolution, SWE-agent studies
agent--computer interfaces, and OpenHands integrates code execution, shell access,
and browsing \citep{jimenez2023swebench,yang2024sweagent,wang2024openhands}.
Terminal-Bench evaluates challenging terminal tasks in task-specific environments
with executable verification; its trajectory analysis distinguishes execution,
coherence, and verification failures \citep{merrill2026terminalbench}. VISTA extends
trajectory analysis to application construction, where process signals such as
design inspection, verification, and completion reports are recorded alongside
component-level outcome scores.

WebArena, VisualWebArena, and OSWorld evaluate agents completing tasks in existing
websites and desktop environments \citep{zhou2023webarena,koh2024visualwebarena,xie2024osworld};
VISTA instead requires agents to construct and launch the application under test.

\section{Benchmark Construction}
\label{sec:benchmark-draft}

\subsection{Task formulation}
\label{sec:task-formulation}

VISTA evaluates application development from multi-page design handoffs.
Each task requires an agent to translate page layouts, interface components,
and interaction requirements into a runnable application. In the full-stack
setting, the interface must integrate with a backend and database to display
data and respond to user actions.

We represent a task as $\tau=(X,E,R)$, where $X$ is the agent-visible
specification, $E$ is the development and execution environment, and $R$
specifies the deliverables. For an application with $P$ pages, the core input
is $X_{\mathrm{core}}=(\{I_p,J_p\}_{p=1}^{P},D)$: $I_p$ is a rendered Figma
page, $J_p$ describes its component hierarchy and geometry, and $D$ provides
textual requirements. The Web track defines six conditions that cross three
input tiers---text requirements only, adding rendered screenshots, and further
adding structure JSON---with a free stack or a task-specific prescribed stack. Under a free
stack these tiers are $C_0$, $C_2$, and $C_4$; the corresponding fixed-stack
conditions are $C_5$, $C_1$, and $C_3$. Main results on both tracks use the full input $X_{\mathrm{core}}$ with a
free stack, which on Web is $C_4$.

The agent acts through the tools of its coding harness (file editing, command
execution, dependency installation, building, and debugging) in unattended
runs without per-call human approval. Web deliverables include application
code and a Docker Compose configuration; Android deliverables include an
installable APK, a build script, and a containerized backend and database.
Web evaluation runs in a browser; Android evaluation uses an emulator accessed
through ADB and Appium.

\subsection{Collection and annotation}

\paragraph{Design selection and export.}
We select multi-page Figma designs that describe complete applications with
connected interfaces, so that tasks require implementing application
workflows. For each design, we export rendered pages and node structures:
images provide visual references, structure files provide hierarchy and
geometry, and descriptions specify component purposes and interactions
\citep[cf.][]{deka2017rico,wang2021screen2words}. Android tasks additionally
provide navigation relations and component-level assets.

\paragraph{Human interaction annotation.}
Figma layer names do not reliably specify behavior: a button may carry a
generic name, and repeated instances may lack explicit action semantics.
We therefore manually annotate interactable components in an annotation
interface, associating their positions with expected behavior
(Figure~\ref{fig:construction-draft}). Records contain
bounding boxes, interaction types, and, where applicable, navigation targets,
behavior subtypes, and semantic identifiers. They cover navigation, input,
toggles, and different click actions. The annotations establish component-level interaction ground truth: reference
geometry identifies where a control belongs, interaction types determine the
behavior to probe, and annotated targets specify navigation destinations where
applicable. Together with task descriptions, they turn static designs into
implementable and testable requirements \citep{xie2007oracles}.

The Web data contain 3,253 interaction annotations and a separate set of
458 curated visual anchors for identifying corresponding elements and
assisting coordinate alignment. The primary score covers the \emph{critical}
subset of annotations: all navigation, input, and toggle components, plus
click components whose annotated action is defined (excluding decorative or
unresolved clicks); remaining annotations contribute no score.

\subsection{Scale and coverage}

\begin{figure}[t]
\centering
\includegraphics[width=\linewidth]{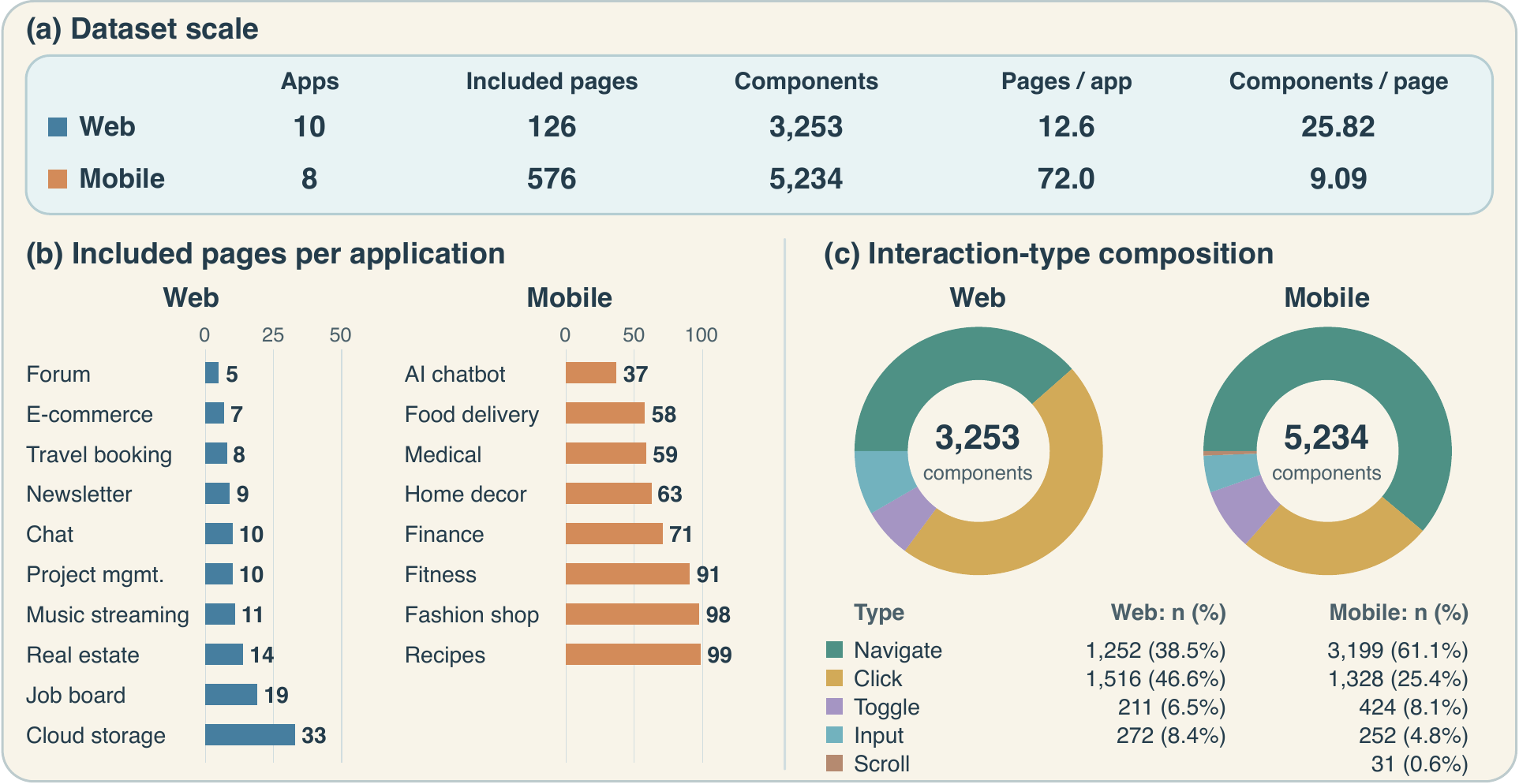}
\caption{Dataset statistics. (a) Scale over included pages.
(b) Per-application page counts, with separate axes for Web (0--50) and
Android/Mobile (0--100). Blue denotes Web and peach denotes Mobile in (a--b).
(c) Interaction-type composition with consistent category colors. Ring centers
show total components; the legend lists counts and percentages of all annotations.
Web's single-record Scroll and Other categories are omitted from the rings and
legend, and displayed sectors are normalized to fill each ring.}
\label{fig:benchmark-statistics}
\end{figure}

VISTA contains ten Web applications and eight Android applications, which form
the Web and Mobile tracks (Figure~\ref{fig:benchmark-statistics}). Web includes all 126 annotated pages;
Android includes 576 of 647 exported screens after excluding near-duplicate
or out-of-scope screens. These pages carry 3,253 annotated interactive
components on Web and 5,234 on Android, 8,487 in total; visual anchors are
counted separately.

Web tasks cover publishing, real estate, recruitment, forums, travel booking,
chat, cloud storage, e-commerce, project management, and music streaming;
Android tasks cover medical appointments, fitness, home-decor shopping, food
delivery, recipes, personal finance, AI chat, and e-commerce. The two
platforms pose different workloads: Android applications average 72.0 screens
versus 12.6 Web pages ($5.7\times$), while Web pages carry more annotated
components (25.8 versus 9.1), and navigation dominates Android annotations
(61.1\%) whereas click actions dominate Web (46.6\%).

\section{Evaluation}
\label{sec:evaluation-draft}

VISTA grounds execution-based evaluation in human-annotated component
requirements. For each selected interaction, the evaluator identifies a
corresponding runtime element, executes the relevant behavior probe, and
records the matching and interaction evidence used for scoring, so that each
component score links to an explicit requirement under the same rules across
generated applications. Web applications are inspected through the DOM using
Playwright; Android applications through the UI hierarchy using ADB and
Appium, with a shared score formulation and platform-specific matching and
probes.

\paragraph{Localization and behavior.}
Localization $L_i\in[0,1]$ measures component correspondence and spatial
alignment using geometry, component types, and available semantic
identifiers. Behavior $B_i\in[0,1]$ measures evidence from executing the
annotated interaction, such as accepting input, changing a toggle state,
opening a dialog, or navigating. The Web evaluator includes
semantic-anchor matching and content-change navigation checks; Android
uses one-to-one component assignment and target-screen fingerprints.
Partial-credit rules distinguish levels of probe evidence. Together, these
metrics assess component realization and observable interaction outcomes \citep{barr2015oracle,xie2007oracles}.

\paragraph{Joint score.}
For a run on task $\tau$, let $C_\tau$ contain its critical interaction
records and $J_\tau\subseteq C_\tau$ contain those with a non-missing behavior
verdict. The joint score is
\begin{equation}
S_\tau=\frac{1}{|J_\tau|}\sum_{i\in J_\tau}L_i B_i.
\label{eq:evaluation-draft-score}
\end{equation}
This is a mean of component-wise products: an interaction receives no joint
credit when either score is zero. Mean localization over $C_\tau$ and mean
behavior over $J_\tau$ are reported separately to distinguish correspondence
from interaction failures. Web uses $J_\tau=C_\tau$; Android excludes
records whose behavior verdict cannot be determined. A delivery that fails to
launch, or a page that cannot be reached, receives $L_i=B_i=0$ for every
affected component and is averaged in with the other runs.

\paragraph{Human validation.}
Blind human studies support both scores (Appendix~\ref{app:human-validation}).
Localization correlates with three annotators' fidelity ratings of generated
Web pages at Spearman $\rho=0.66$ (95\% CI $[0.36,0.86]$; Krippendorff's
$\alpha=0.70$), whereas CLIP image similarity does not ($\rho=0.04$). Behavior
probe verdicts agree with human inspection of before-and-after screenshots in
88\% of cases (Cohen's $\kappa=0.68$).

\paragraph{Results.}
Table~\ref{tab:evaluation-web-mobile} lists Web and Mobile results by
model--harness configuration, with separate columns for localization $L$,
behavior $B$, joint score $S$, and mean token usage per task for 14
model--harness configurations. Each system runs every task three times, and
all values are means over runs. Two patterns stand out. First, behavior trails
localization: 13 of 14 systems score lower on $B$ than on $L$ on Web (mean
0.543 versus 0.642), so placing a control is consistently easier than making it
work. Second, rankings diverge across platforms and metrics: GPT-6-astra ranks seventh on Web $S$ but first on
Mobile, while the Web leader, Grok 4.6, ranks fifth on Mobile; on Web,
GPT-6-astra leads localization ($L=0.756$) and Grok 4.6 leads behavior
($B=0.664$). A single-platform or single-metric leaderboard would miss these
differences. Both tracks leave substantial headroom: the best joint score is
0.553 on Web and 0.378 on Mobile.
\begin{table}[t]
\centering
\providecommand{\vistaModelIcon}[2]{\makebox[1.35em][c]{\raisebox{-0.14em}{\includegraphics[height=0.95em,trim=#2,clip]{figs/model_icons/#1.png}}}\hspace{0.25em}}
\providecommand{\vistaNoModelIcon}{\makebox[1.35em][c]{}\hspace{0.25em}}
\footnotesize
\setlength{\tabcolsep}{2pt}
\renewcommand{\arraystretch}{1.08}
\caption{Web and Mobile evaluation results. Both tracks use the full input with a free implementation stack ($C_4$). Localization $L$, behavior $B$, and joint score $S$ are reported separately; higher is better. Token usage $T$ is the mean non-cached (input plus output) tokens per task in thousands; Appendix~\ref{app:token-usage} breaks usage down by token type. Each value is the mean over three runs per task. Overall is the mean of Web and Mobile $S$; rows are ordered by Overall. Model sources and harness versions are listed in Appendix~\ref{app:model-sources}.}
\label{tab:evaluation-web-mobile}
\begin{tabular*}{\linewidth}{@{\extracolsep{\fill}}llccccccccc@{}}
\toprule
Model & Harness & \multicolumn{4}{c}{Web} & \multicolumn{4}{c}{Mobile} & Overall \\
\cmidrule(lr){3-6}\cmidrule(lr){7-10}
 & & $L$ & $B$ & $S$ & $T$ & $L$ & $B$ & $S$ & $T$ & $\uparrow$ \\
\midrule
\vistaModelIcon{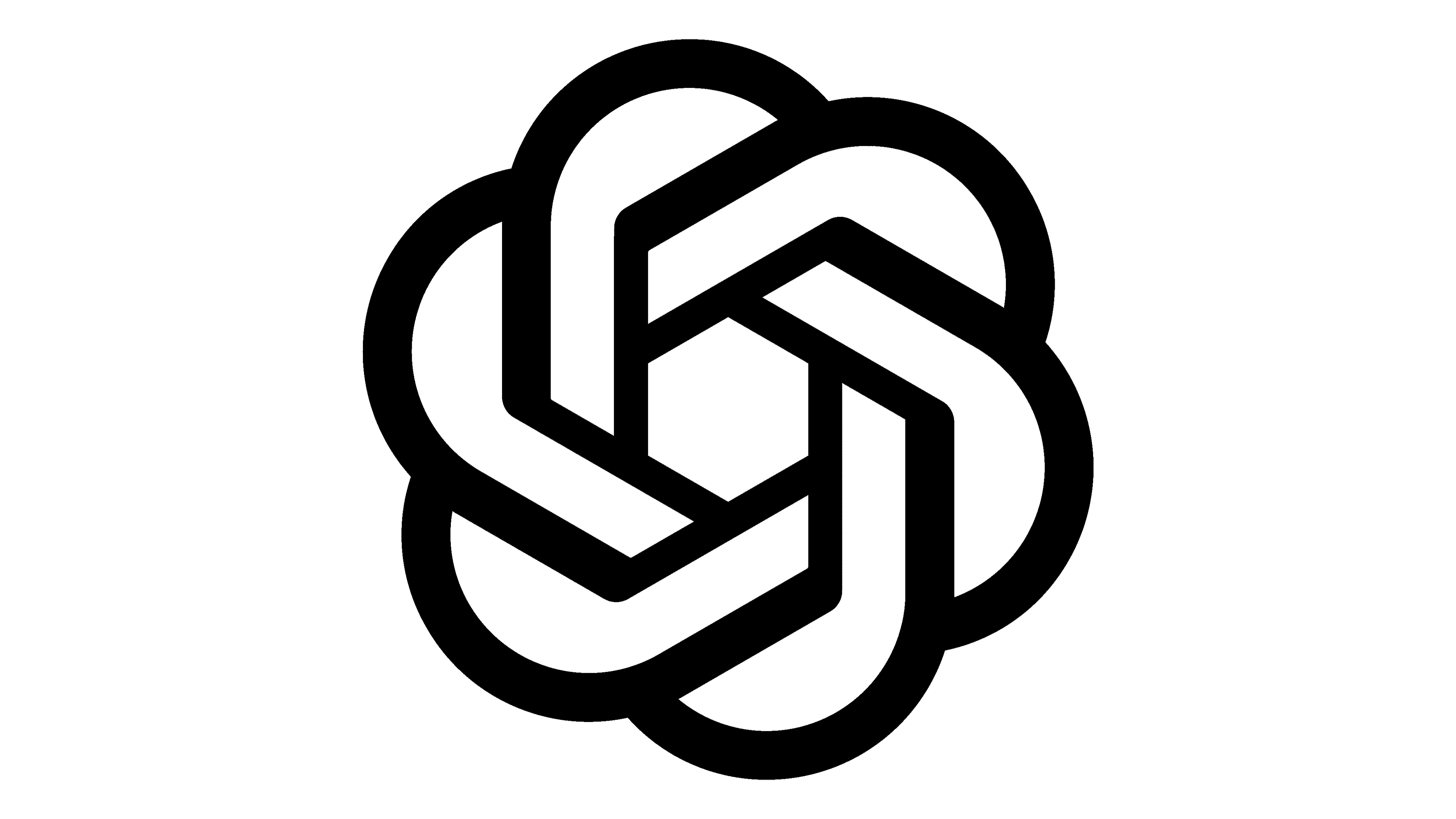}{956 102 956 102}GPT-6-astra & Codex & $0.756$ & $0.505$ & $0.460$ & 180 & $0.473$ & $0.398$ & $0.378$ & 387 & $0.419$ \\
\vistaModelIcon{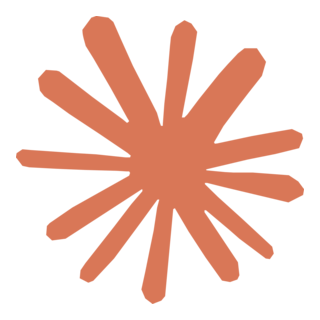}{16 16 16 16}Fable-5 & Claude Code & $0.749$ & $0.609$ & $0.533$ & 366 & $0.367$ & $0.402$ & $0.270$ & 1020 & $0.402$ \\
\vistaModelIcon{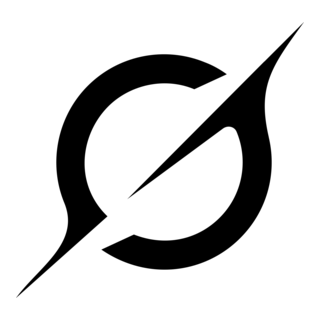}{16 22 16 22}Grok 4.6 & Cursor & $0.732$ & $0.664$ & $0.553$ & 444 & $0.336$ & $0.379$ & $0.234$ & 387 & $0.393$ \\
\vistaModelIcon{claude}{16 16 16 16}Opus 4.8 & Claude Code & $0.686$ & $0.572$ & $0.487$ & 291 & $0.395$ & $0.424$ & $0.282$ & 565 & $0.385$ \\
\vistaModelIcon{grok}{16 22 16 22}Grok 4.5 & Cursor & $0.670$ & $0.590$ & $0.491$ & 154 & $0.418$ & $0.353$ & $0.245$ & 411 & $0.368$ \\
\vistaModelIcon{openai}{956 102 956 102}GPT-5.6-sol & Codex & $0.738$ & $0.593$ & $0.507$ & 351 & $0.377$ & $0.272$ & $0.228$ & 99 & $0.367$ \\
\vistaModelIcon{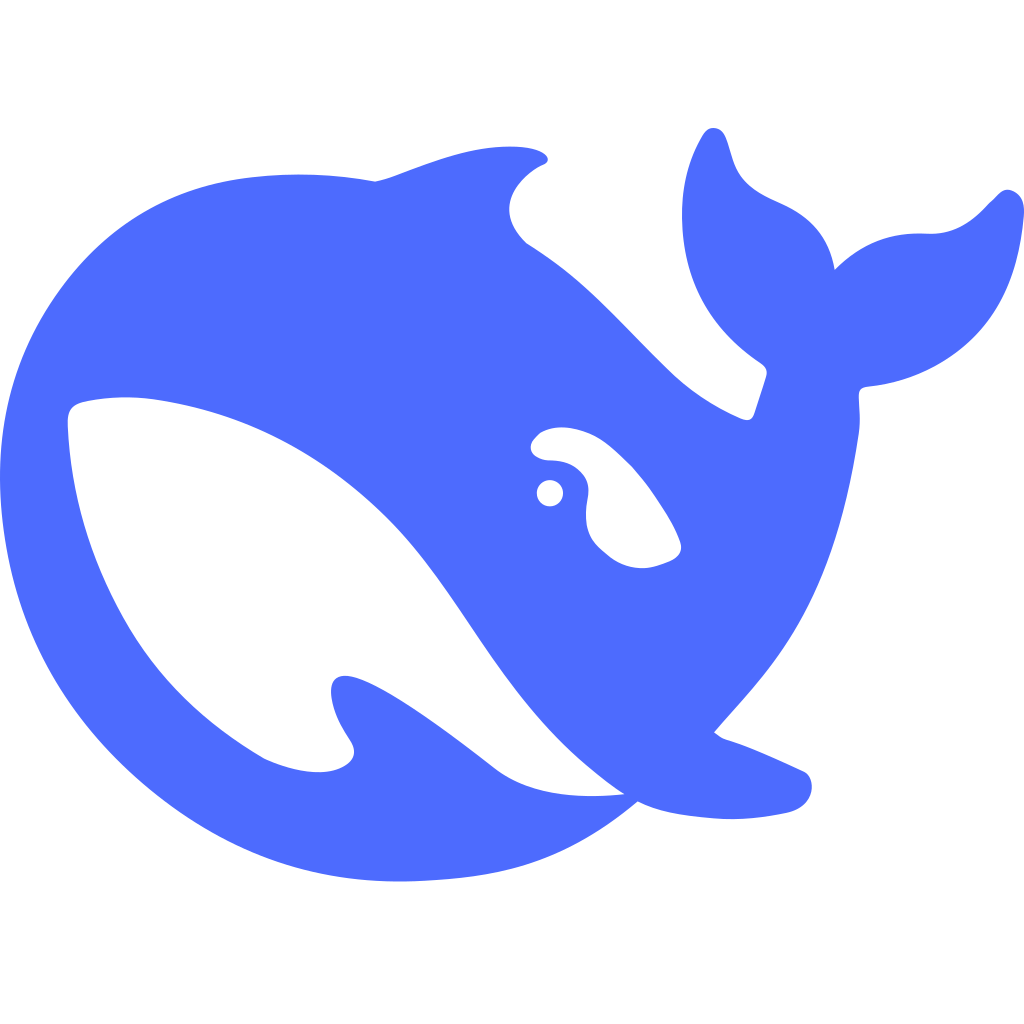}{0 142 0 128}DeepSeek-V4 & DeepSeek CLI & $0.714$ & $0.581$ & $0.484$ & 360 & $0.401$ & $0.275$ & $0.219$ & 373 & $0.351$ \\
\vistaModelIcon{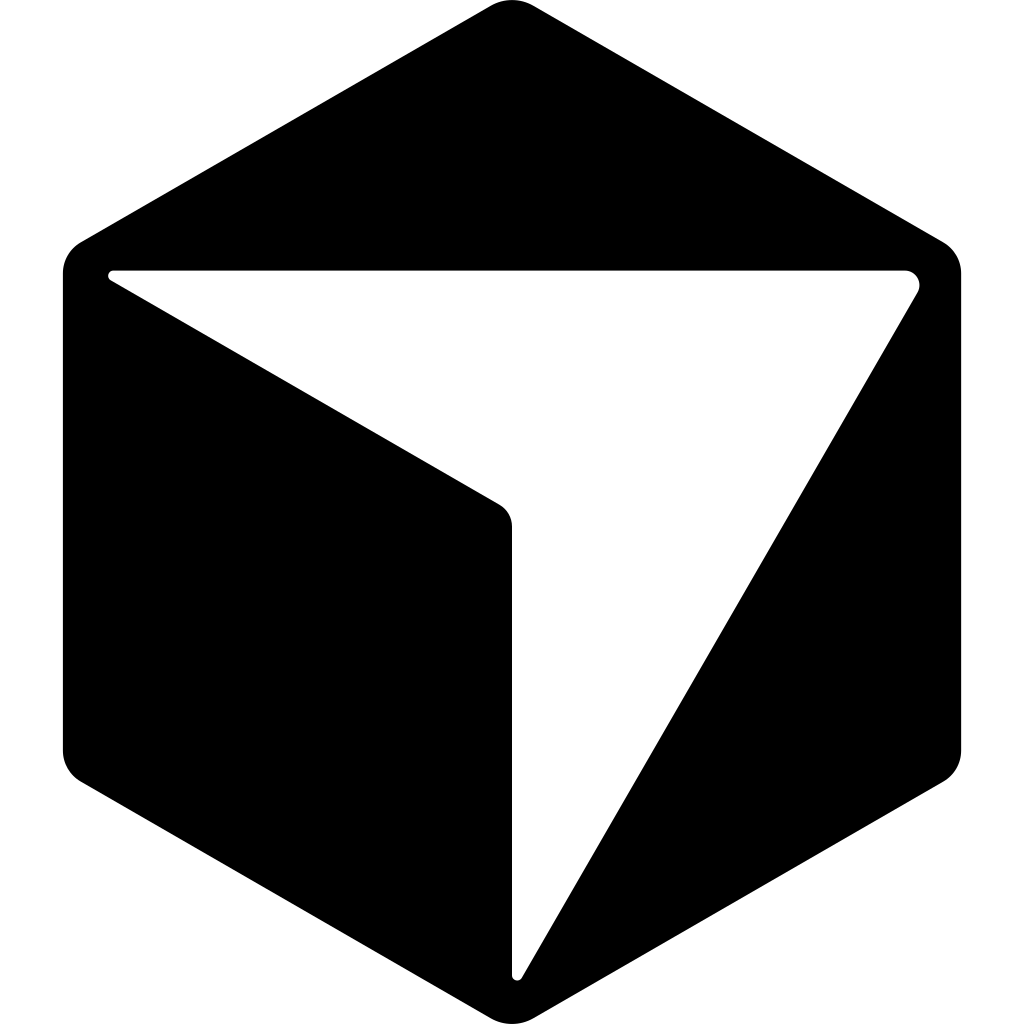}{62 0 62 0}Composer 2.5 & Cursor & $0.599$ & $0.569$ & $0.448$ & 116 & $0.413$ & $0.300$ & $0.192$ & 122 & $0.320$ \\
\vistaModelIcon{openai}{956 102 956 102}GPT-5.5 & Codex & $0.656$ & $0.478$ & $0.410$ & 253 & $0.310$ & $0.287$ & $0.125$ & 199 & $0.268$ \\
\vistaModelIcon{claude}{16 16 16 16}Sonnet 4.6 & Claude Code & $0.586$ & $0.517$ & $0.412$ & 150 & $0.265$ & $0.205$ & $0.115$ & 301 & $0.264$ \\
\vistaModelIcon{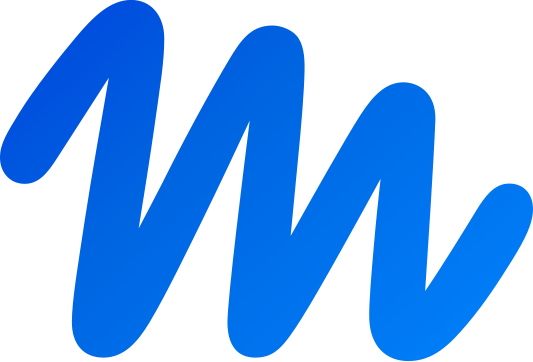}{0 0 0 0}Muse-Spark-1.2 & Muse Code & $0.546$ & $0.542$ & $0.409$ & 180 & $0.287$ & $0.294$ & $0.110$ & 369 & $0.259$ \\
\vistaModelIcon{openai}{956 102 956 102}GPT-5.6-terra & Codex & $0.570$ & $0.488$ & $0.384$ & 75 & $0.246$ & $0.198$ & $0.072$ & 92 & $0.228$ \\

\vistaModelIcon{openai}{956 102 956 102}GPT-5.6-luna & Codex & $0.581$ & $0.482$ & $0.387$ & 84 & $0.113$ & $0.121$ & $0.028$ & 101 & $0.207$ \\
\makebox[1.35em][c]{\raisebox{-0.05em}{\includegraphics[width=1.3em,trim=23 141 22 140,clip]{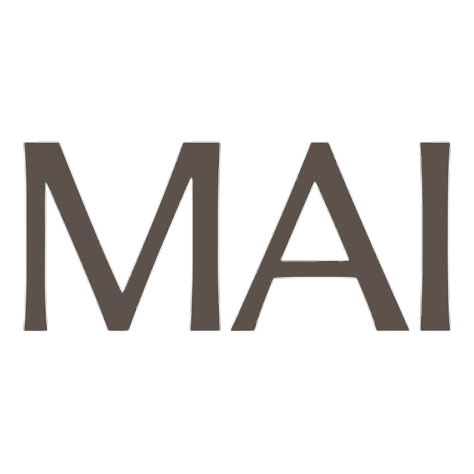}}}\hspace{0.25em}MAI-Code-1.1 & Copilot & $0.409$ & $0.418$ & $0.293$ & 153 & $0.220$ & $0.137$ & $0.055$ & 234 & $0.174$ \\
\bottomrule
\end{tabular*}
\vspace{3pt}
\begin{minipage}{\linewidth}
\footnotesize\raggedright
\textit{Note.} Muse-Spark-1.2 denotes Muse-Spark-1.2-contributor;
DeepSeek-V4 denotes DeepSeek-V4-Flash; MAI-Code-1.1 denotes MAI-Code-1.1-Flash.
Tokens exclude cache-read.
\end{minipage}
\end{table}

\FloatBarrier
\section{Agent Workflow Analysis}
\label{sec:workflow}

Component-level scores show that delivered applications fall short of their
designs; development traces show where the process departs from what the task
requires.

\subsection{Failure modes in delivered applications}
\label{sec:failure-examples}

\paragraph{Unresponsive controls are widespread across Web deliveries.}
In a Web audit of 12 systems , 90.1\% of scored deliveries contain at least one control whose
interaction probe detects no response, and 13.6\% receive a joint score below
0.2. Rendering a control and implementing its behavior are
separate achievements, and component-level execution exposes the difference.

\paragraph{Mobile failures extend to deployment.}
On Mobile, the failure spectrum begins before any interaction can be probed:
2.9\% of backend-evaluation deliveries mount an empty directory as
\texttt{init.sql}, preventing database initialization, and others ship
incomplete schemas, invalid SQL comment syntax, mismatched \texttt{INSERT}
values, missing Dockerfiles, or unlaunchable Android Activities. Such defects break the path from
generated code to a running application, a stage that screenshot-based
evaluation cannot observe.

\subsection{Alignment gaps in design grounding, verification, and reporting}
\label{sec:alignment}

\begin{figure}[t]
\centering
\includegraphics[width=\linewidth]{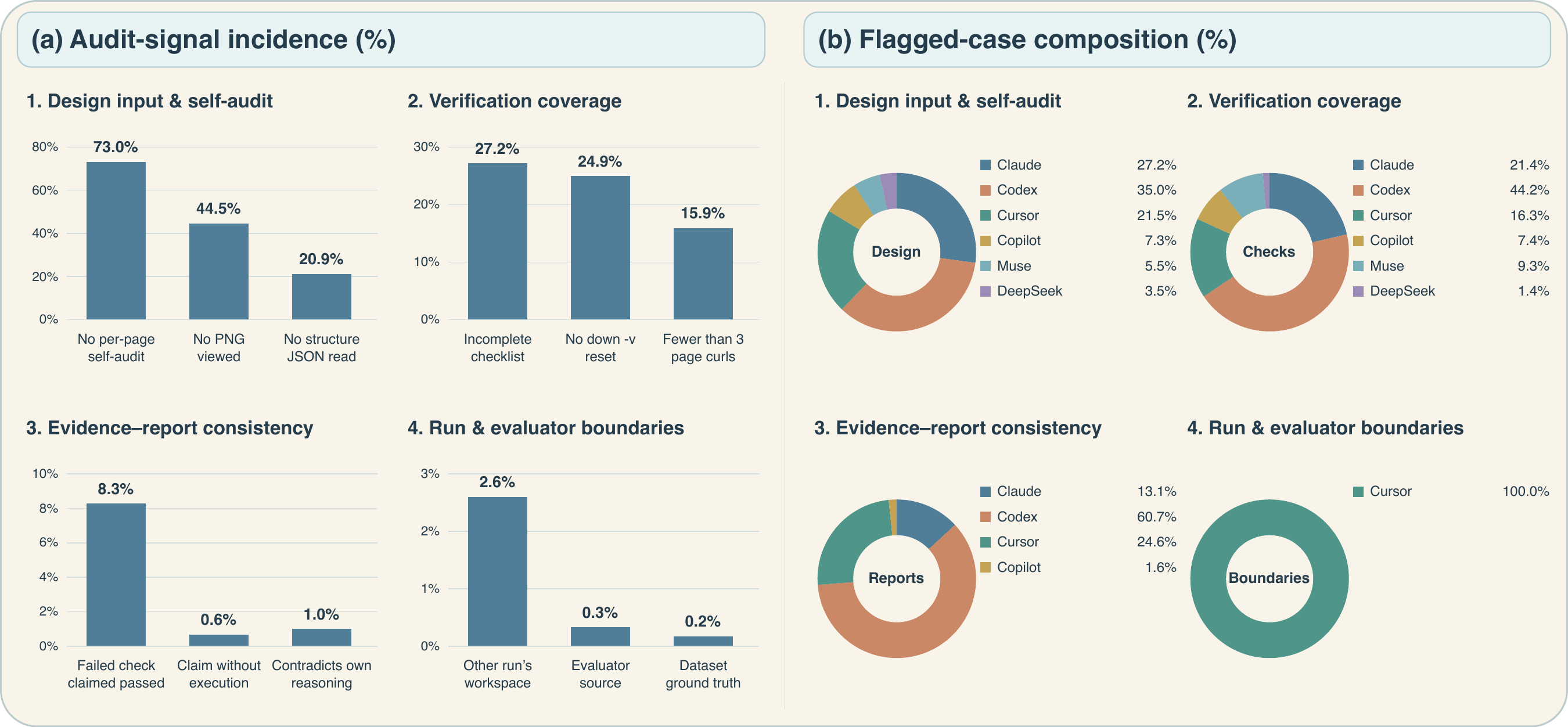}
\caption{Alignment gaps across four directions. Left: detection rates as percentages
of all audited trajectories; axis ranges vary by panel. Right: system-group shares
within each direction's flagged cases, with each trajectory counted once per
direction. Report--reasoning contradictions require visible reasoning and are
reported using the same all-trajectory denominator.}
\label{fig:alignment-signals}
\end{figure}

Development traces reveal systematic gaps between what the task prescribes and
what agents do (Figure~\ref{fig:alignment-signals}); stack compliance is
measured in the fixed-stack conditions of the condition study.

\paragraph{Agents bypass the design evidence they are given.}
The system prompt instructs agents to read the screenshot, structure JSON, and
description before writing each page, and to self-audit each page against them
(Appendix~\ref{app:system-prompt-web}). Yet no screenshot inspection is
recorded in 44.5\% of trajectories, no structure JSON read in 20.9\%, and the
per-page self-audit is absent in 73.0\%. The omission reflects behavior rather
than capability: Codex supports image input, yet 71.1\% of its trajectories
never open a screenshot, compared with 21.8\% for Claude Code and Cursor.

\paragraph{Stack mandates are followed; process instructions are not.}
In the fixed-stack conditions, each task prescribes a frontend framework such
as Astro, Next.js, or SvelteKit. For GPT-5.5 and Sonnet 4.6, the detected
framework matches the prescription in 88.3\% of runs across $C_1$, $C_3$, and
$C_5$; the remainder, all by GPT-5.5, are detected as React where Next.js was
prescribed. Agents thus follow instructions about what to build with, yet skip
the prescribed steps for checking what they build against the design.

\paragraph{Success claims outrun execution evidence.}
The audit flags incomplete verification in 27.2\% of trajectories and a failed
check followed by a claim that it passed in 8.3\%. In a job-board trace, a
\texttt{set -e} verification block terminates when \texttt{wget} is unavailable,
yet the closing message reports successful login and page checks. Agent
self-reports are therefore not a substitute for component-level execution
checks.

\subsection{Richer inputs increase activity without measurable score gains}
\label{sec:input-stack-analysis}

For GPT-5.5 and Sonnet 4.6, the two models covered by the condition study,
richer design inputs increase agent activity: under a free stack, tool use rises from 53.5 calls with text alone
($C_0$) to 69.3 with screenshots ($C_2$) and 71.5 with additional structure
JSON ($C_4$); token workload is 134.7K, 224.1K, and 149.3K, respectively.
Joint scores do not follow: across the four input and stack contrasts,
estimated changes stay within $\pm0.023$, with all 95\% confidence intervals
spanning zero (Appendix~\ref{app:condition-contrasts}). Agents do more with
richer inputs without measurably converting the additional design evidence into
higher fidelity, consistent with the frequent omission of screenshot inspection in
Section~\ref{sec:alignment}.

\subsection{Workflow signatures across harnesses}
\label{sec:workflow-signatures}

\begin{figure}[t]
    \centering
    \includegraphics[width=\linewidth]{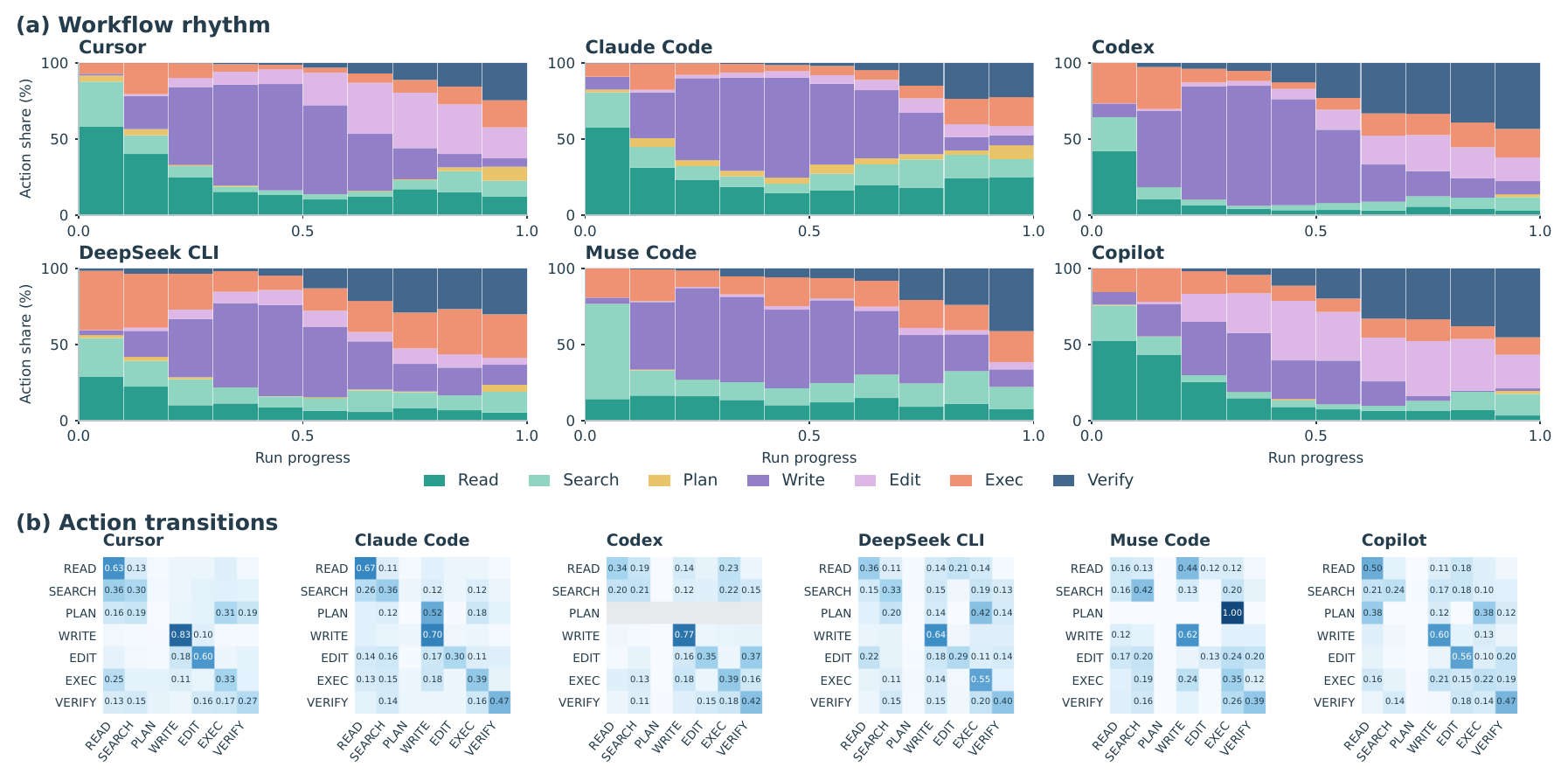}
    \caption{Workflow signatures across six harness groups covering 14 models. (a) Action shares over ten normalized progress bins, arranged
    in two rows; category shares are normalized within each model/bin before equal-weight
    averaging across models. Pale vertical lines mark 0.1 progress intervals.
    (b) Transition probabilities pooled across observed transitions and normalized
    by the source action. Both panels use the same seven action categories, including VERIFY. Cells below 0.10
    are unannotated; gray rows have no observed source transitions.
    Complete model profiles appear in Appendix~\ref{app:workflow-trajectories}.}
    \label{fig:workflow-trajectories}
\end{figure}

\paragraph{Agents shift from inspection to construction to verification.}
Across the six harness groups (Figure~\ref{fig:workflow-trajectories}), READ
and SEARCH occupy 41.3--70.1\% of activity in the first two progress bins,
WRITE and EDIT 53.0--76.3\% in bins four through seven, and VERIFY rises from
0.1--2.6\% in the first two bins to 20.0--41.6\% in the last two.

\paragraph{Harnesses differ in when they verify.}
EDIT-to-VERIFY transitions account for 37.2\% in Codex, compared with 7.5\% in
Cursor and 9.5\% in Claude Code; Cursor and Claude Code instead show strong
READ persistence (62.7\% and 67.2\% READ-to-READ, versus 33.8\% in Codex), and
Cursor the strongest EDIT persistence (60.2\%).

\begin{figure}[t]
    \centering
    \includegraphics[width=\linewidth]{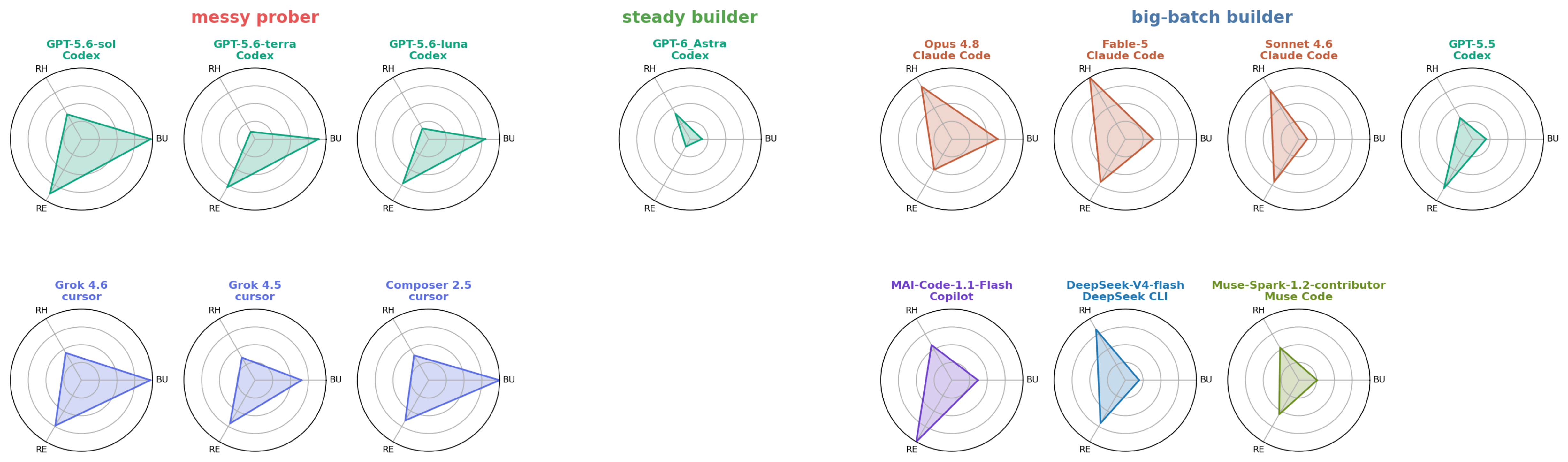}
    \caption{Behavioral profiles of 14 models in three clusters: messy prober
    (six models), steady builder (one), and big-batch builder (seven).
    BU (\texttt{build\_share}) is the fraction of actions that are WRITE or EDIT.
    RH (\texttt{gap\_cv}) is the coefficient of variation of the gaps between
    build bursts, where a burst is a maximal run of consecutive WRITE or EDIT
    actions and a gap is the number of other actions between two bursts; low
    values indicate regular pacing and high values indicate clustered activity. RE (\texttt{rework\_rate}) is the fraction of touched files revisited
    more than once. Colors identify harness groups.}
    \label{fig:behavioral-profiles}
\end{figure}

\paragraph{Workflow style varies with the model within a harness.}
Figure~\ref{fig:behavioral-profiles} characterizes 14 models through build share
(BU), rhythm (RH), and rework (RE). GPT-6-astra forms a singleton ``steady
builder'' cluster with low rework and regular pacing, interleaving writing and
execution (WRITE-to-EXEC 26.5\%, versus 0.7--3.8\% for the other Codex models;
Appendix~\ref{app:workflow-trajectories}). The ``messy prober'' cluster
(GPT-5.6-sol, -terra, -luna, and the three Cursor models) pairs prominent build
shares with frequent revisitation, and the ``big-batch builder'' cluster (the
three Claude Code models, GPT-5.5, DeepSeek, Muse, and Copilot) shows pronounced
gap variability. Codex models span all three clusters, so workflow style varies
with the model even within a shared harness.

\section{Discussion}
\label{sec:discussion}

The divergence between localization and behavior motivates agents that track
both the placement of each required component and the interaction it must
support, and the condition study suggests that the bottleneck lies in using
design evidence rather than accessing it: richer inputs increase tool activity
without a measurable change in joint score. Connecting design inspection to
explicit component requirements, and tying completion reports to
component-level execution evidence, are directions VISTA can measure directly.

\paragraph{Limitations.}
Our results characterize deployed model--harness configurations jointly,
without isolating model and harness effects, and the input and stack contrasts
cover two models. Tasks use self-contained backends; authentication,
permissions, and external APIs are natural extensions of the same methodology.

\section{Conclusion}
\label{sec:conclusion}

VISTA evaluates coding agents on the full path from a visual design to a
running full-stack application on Web and Mobile. Its component-level checks
show that current agents place designs more reliably than they implement them,
and its development traces show agents bypassing the design evidence and
self-checks they are given while reporting success beyond what their checks
support. VISTA provides the scores and traces needed to measure progress on
closing this gap.

\section*{Disclosure of AI Use}

In this work, we used generative AI tools to assist with implementing
methods, translating text, and creating and editing software code.
The authors take responsibility for the final content of this work,
including all text, claims, and artifacts produced with the assistance
of generative AI.

\bibliographystyle{plainnat}
\bibliography{references}

\appendix
\clearpage
\section{Model Sources}
\label{app:model-sources}
\begin{table}[h]
\centering
\small
\renewcommand{\arraystretch}{1.15}
\caption{Official sources and harness versions for the systems evaluated in Table~\ref{tab:evaluation-web-mobile}. All runs of a harness use the same pinned version.}
\label{tab:model-sources}
\begin{tabular*}{\linewidth}{@{\extracolsep{\fill}}llll@{}}
\toprule
Model & Source & Harness & Version \\
\midrule
GPT-6-astra & \citet{openai2026gpt6astra} & Codex & 0.144.6 \\
Fable-5 & \citet{anthropic2026fable5} & Claude Code & 2.1.187 \\
Grok 4.6 & \citet{spacexai2026grok46} & Cursor & 2026.09.10 \\
Opus 4.8 & \citet{anthropic2026opus48} & Claude Code & 2.1.187 \\
Grok 4.5 & \citet{spacexai2026grok45} & Cursor & 2026.09.10 \\
GPT-5.6-sol & \citet{openai2026gpt56sol} & Codex & 0.144.6 \\
DeepSeek-V4 & \citet{deepseek2026v4flash} & DeepSeek CLI & 0.1.5 \\
Composer 2.5 & \citet{cursor2026composer25} & Cursor & 2026.09.10 \\
GPT-5.5 & \citet{openai2026gpt55} & Codex & 0.144.6 \\
Sonnet 4.6 & \citet{anthropic2026sonnet46} & Claude Code & 2.1.187 \\
Muse-Spark-1.2 & \citet{reso2026musecode} & Muse Code & 1.3.0 \\
GPT-5.6-terra & \citet{openai2026gpt56terra} & Codex & 0.144.6 \\
GPT-5.6-luna & \citet{openai2026gpt56luna} & Codex & 0.144.6 \\
MAI-Code-1.1 & \citet{microsoft2026maicode1flash} & Copilot & 1.0.86 \\
\bottomrule
\end{tabular*}
\end{table}

\clearpage
\section{Token Usage by Type}
\label{app:token-usage}
\begin{figure}[h]
\centering
\includegraphics[width=\linewidth]{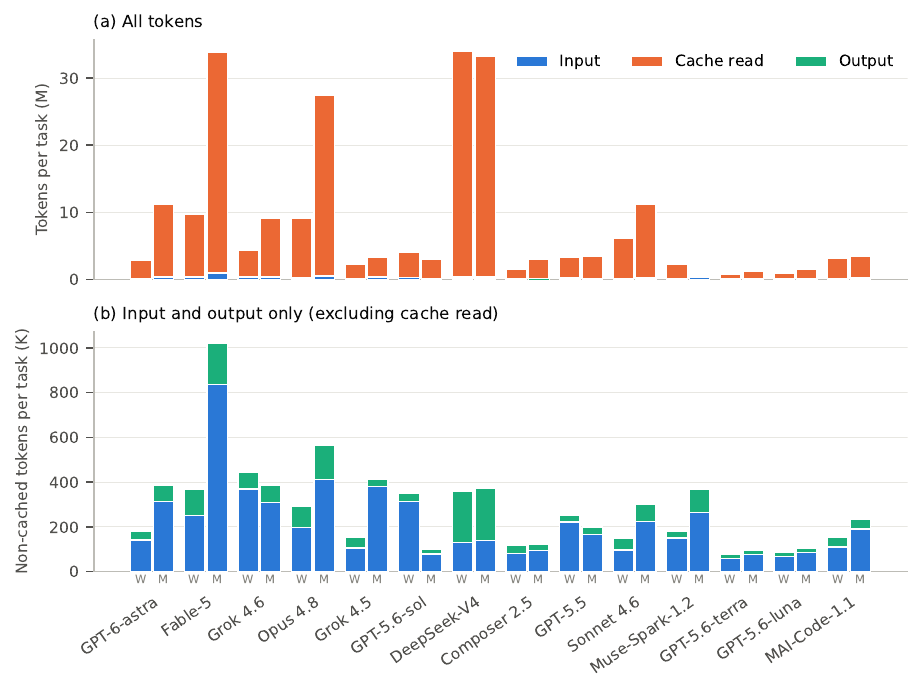}
\caption{Mean per-task token usage by type. Each model has a Web (W, darker
shades) and a Mobile (M, lighter shades) bar. (a) All tokens, in millions. (b) Input and output tokens only,
in thousands, corresponding to $T$ in Table~\ref{tab:evaluation-web-mobile}.}
\label{fig:token-usage}
\end{figure}

Figure~\ref{fig:token-usage} breaks down mean per-task token usage into
input, cache-read, and output tokens for each system on Web (W) and Mobile (M).
Cache reads account for 87--99\% of all tokens in every system--platform
pair except Muse-Spark-1.2 on Mobile (41\%). Table~\ref{tab:evaluation-web-mobile}
reports the non-cached total (input plus output), shown in panel (b).

\clearpage
\section{Human Validation of the Evaluator}
\label{app:human-validation}

\paragraph{Localization.}
Three annotators independently rated the visual and structural fidelity of
generated pages against the reference designs on a five-point Likert scale,
following a written rubric. Items were presented in randomized order without
model or condition labels. The study covers 40 generated Web applications
(118 pages) from four agent systems under the screenshot conditions $C_2$ and
$C_4$. Localization correlates with the mean human rating at Spearman
$\rho=+0.66$ (95\% CI $[+0.36,+0.86]$, $n=40$), and the joint score at
$\rho=+0.43$ ($[+0.08,+0.70]$). Inter-annotator agreement is Krippendorff's
$\alpha=0.70$ (ordinal), with a mean pairwise Spearman correlation of 0.76 and
96\% of rating pairs within one point. At the system level, Kendall's
$\tau=+0.33$ between human and automatic rankings; with four closely scoring
systems this is a coarse check, and the item-level correlation is the primary
evidence.

\paragraph{Comparison with CLIP.}
CLIP image similarity does not track the same human judgments
($\rho=+0.04$, $[-0.49,+0.58]$, $n=20$). CLIP measures holistic image
appearance, including photographs and other image content that VISTA does not
require agents to reproduce; agents routinely substitute their own placeholder
images. VISTA therefore scores structural fidelity through DOM-grounded
component matching.

\paragraph{Behavior probes.}
Three raters inspected before-and-after interaction screenshots for 24 probed
behavior anchors, with 92\% inter-rater agreement. The inspection showed that a
URL-based navigation criterion misjudges single-page applications in both
directions: content can change without a URL change, and a URL can change while
the page renders a fallback. The evaluator therefore uses a content-based
criterion, under which a navigation or pop-out interaction succeeds when the
click produces the corresponding observable content change. Under this
criterion, probe verdicts agree with human judgment in 88\% of cases
(Cohen's $\kappa=0.68$), compared with 50\% ($\kappa=-0.20$) under the
URL-based criterion. Remaining disagreements concern input-value edge cases.

\clearpage
\section{Agent System Prompts}
\label{app:system-prompts}
We reproduce the system prompts for the Web and Mobile tracks below.
Formatting is adapted for readability; instructions, commands, and examples
are preserved. Task-specific descriptions are appended to these prompts
when agents are run.

\subsection{Web Application Generation}
\label{app:system-prompt-web}
\begin{SystemPromptBox}{Web System Prompt}
\PromptSourceTitle{Agent System Prompt --- Web App Code Generation \& Deployment (variant \textbf{c4})}
\textbf{Difference vs {\promptfont a\allowbreak{}g\allowbreak{}e\allowbreak{}n\allowbreak{}t\allowbreak{}\_\allowbreak{}s\allowbreak{}y\allowbreak{}s\allowbreak{}t\allowbreak{}e\allowbreak{}m\allowbreak{}\_\allowbreak{}p\allowbreak{}r\allowbreak{}o\allowbreak{}m\allowbreak{}t\allowbreak{}\_\allowbreak{}c\allowbreak{}3\allowbreak{}.\allowbreak{}m\allowbreak{}d}:} The {\promptfont [\allowbreak{}S\allowbreak{}C\allowbreak{}A\allowbreak{}F\allowbreak{}F\allowbreak{}O\allowbreak{}L\allowbreak{}D\allowbreak{}I\allowbreak{}N\allowbreak{}G\allowbreak{}]} section is rewritten --- \textbf{no scaffold command is provided in the task description for c4}. Pick a framework and bootstrap method yourself, based on what best fits the mockup and described behavior. All other constraints (visual fidelity, testid contract, verification, etc.) are unchanged.

\PromptHeading{ROLE}
Senior full-stack + DevOps engineer. Full read/write/execute in the project folder. Use tools and skills freely; act, verify, fix --- do not ask for permission. Output must be a working app the evaluator can spin up with one command and test without any manual step.

\textbf{You are running fully unattended in a non-interactive batch.} There is NO user available to answer questions. Never present options ("Option A vs Option B"), never ask "which framework do you prefer", never wait for confirmation. When the task description leaves a choice open (framework, library, styling approach, database, etc.), pick the most reasonable option \textbf{silently} and proceed. Treat every ambiguity as your call to make. The only valid termination is a working app that meets [VERIFICATION] --- anything that ends with a question to the user counts as a failed run.

\PromptHeading{SCAFFOLDING --- your call to make}
The task description does NOT prescribe a framework or scaffold command. \textbf{You pick the stack} based on the task's nature:

\begin{enumerate}
\item \textbf{Inspect the mockup and description first.} A blog-style content site might fit Astro / Eleventy; a heavy interactive dashboard might fit Next.js / Remix; a Q\&A or chat app might fit SvelteKit; an e-commerce template might fit Next.js + Stripe; a Django + HTMX SSR setup is fine for content-heavy sites with simple interactions.
\item \textbf{Use a real scaffold} rather than hand-writing configs from scratch. {\promptfont n\allowbreak{}p\allowbreak{}x\allowbreak{} \allowbreak{}c\allowbreak{}r\allowbreak{}e\allowbreak{}a\allowbreak{}t\allowbreak{}e\allowbreak{}-\allowbreak{}n\allowbreak{}e\allowbreak{}x\allowbreak{}t\allowbreak{}-\allowbreak{}a\allowbreak{}p\allowbreak{}p\allowbreak{}@\allowbreak{}l\allowbreak{}a\allowbreak{}t\allowbreak{}e\allowbreak{}s\allowbreak{}t}, {\promptfont n\allowbreak{}p\allowbreak{}m\allowbreak{} \allowbreak{}c\allowbreak{}r\allowbreak{}e\allowbreak{}a\allowbreak{}t\allowbreak{}e\allowbreak{} \allowbreak{}v\allowbreak{}i\allowbreak{}t\allowbreak{}e\allowbreak{}@\allowbreak{}l\allowbreak{}a\allowbreak{}t\allowbreak{}e\allowbreak{}s\allowbreak{}t}, {\promptfont n\allowbreak{}p\allowbreak{}m\allowbreak{} \allowbreak{}c\allowbreak{}r\allowbreak{}e\allowbreak{}a\allowbreak{}t\allowbreak{}e\allowbreak{} \allowbreak{}a\allowbreak{}s\allowbreak{}t\allowbreak{}r\allowbreak{}o\allowbreak{}@\allowbreak{}l\allowbreak{}a\allowbreak{}t\allowbreak{}e\allowbreak{}s\allowbreak{}t}, {\promptfont n\allowbreak{}p\allowbreak{}x\allowbreak{} \allowbreak{}s\allowbreak{}v\allowbreak{} \allowbreak{}c\allowbreak{}r\allowbreak{}e\allowbreak{}a\allowbreak{}t\allowbreak{}e}, {\promptfont n\allowbreak{}p\allowbreak{}x\allowbreak{} \allowbreak{}n\allowbreak{}u\allowbreak{}x\allowbreak{}i\allowbreak{}@\allowbreak{}l\allowbreak{}a\allowbreak{}t\allowbreak{}e\allowbreak{}s\allowbreak{}t\allowbreak{} \allowbreak{}i\allowbreak{}n\allowbreak{}i\allowbreak{}t}, {\promptfont d\allowbreak{}j\allowbreak{}a\allowbreak{}n\allowbreak{}g\allowbreak{}o\allowbreak{}-\allowbreak{}a\allowbreak{}d\allowbreak{}m\allowbreak{}i\allowbreak{}n\allowbreak{} \allowbreak{}s\allowbreak{}t\allowbreak{}a\allowbreak{}r\allowbreak{}t\allowbreak{}p\allowbreak{}r\allowbreak{}o\allowbreak{}j\allowbreak{}e\allowbreak{}c\allowbreak{}t}, etc. --- all valid choices. Run the scaffold non-interactively (use {\promptfont -\allowbreak{}-\allowbreak{}n\allowbreak{}o\allowbreak{}-\allowbreak{}i\allowbreak{}n\allowbreak{}s\allowbreak{}t\allowbreak{}a\allowbreak{}l\allowbreak{}l}, {\promptfont -\allowbreak{}-\allowbreak{}n\allowbreak{}o\allowbreak{}-\allowbreak{}g\allowbreak{}i\allowbreak{}t}, {\promptfont -\allowbreak{}-\allowbreak{}t\allowbreak{}s}, {\promptfont -\allowbreak{}-\allowbreak{}y\allowbreak{}e\allowbreak{}s} flags as appropriate to avoid prompts).
\item \textbf{After scaffolding}, install task-specific deps and build pages / components / API routes on top.
\item \textbf{Document your stack choice in the README} (in the {\promptfont \#\allowbreak{}\#\allowbreak{} \allowbreak{}Q\allowbreak{}u\allowbreak{}i\allowbreak{}c\allowbreak{}k\allowbreak{} \allowbreak{}S\allowbreak{}t\allowbreak{}a\allowbreak{}r\allowbreak{}t} and {\promptfont \#\allowbreak{}\#\allowbreak{} \allowbreak{}P\allowbreak{}o\allowbreak{}r\allowbreak{}t\allowbreak{}s} sections).
\item \textbf{Cross-platform}: anything you generate must run inside the Docker containers you'll write. Avoid stack choices that require the dev host's exact CPU arch / OS (e.g., Prisma's binaryTargets must include {\promptfont l\allowbreak{}i\allowbreak{}n\allowbreak{}u\allowbreak{}x\allowbreak{}-\allowbreak{}a\allowbreak{}r\allowbreak{}m\allowbreak{}6\allowbreak{}4\allowbreak{}-\allowbreak{}o\allowbreak{}p\allowbreak{}e\allowbreak{}n\allowbreak{}s\allowbreak{}s\allowbreak{}l\allowbreak{}-\allowbreak{}3\allowbreak{}.\allowbreak{}0\allowbreak{}.\allowbreak{}x} etc. or generate inside the Dockerfile build).
\end{enumerate}
There is no "one correct stack" for any task --- you're graded on whether the result is a working, faithfully-styled app, not on framework choice.

\PromptHeading{CONSTRAINTS}
\PromptSubheading{Auth, DB, Docker}
\begin{itemize}
\item If login required: pre-seed \textbf{exactly one} account (e.g., {\promptfont a\allowbreak{}d\allowbreak{}m\allowbreak{}i\allowbreak{}n\allowbreak{}@\allowbreak{}t\allowbreak{}e\allowbreak{}s\allowbreak{}t\allowbreak{}.\allowbreak{}c\allowbreak{}o\allowbreak{}m} / {\promptfont A\allowbreak{}d\allowbreak{}m\allowbreak{}i\allowbreak{}n\allowbreak{}1\allowbreak{}2\allowbreak{}3\allowbreak{}4\allowbreak{}!}) via init script or migration, document in README.
\item Database: local Docker service in {\promptfont d\allowbreak{}o\allowbreak{}c\allowbreak{}k\allowbreak{}e\allowbreak{}r\allowbreak{}-\allowbreak{}c\allowbreak{}o\allowbreak{}m\allowbreak{}p\allowbreak{}o\allowbreak{}s\allowbreak{}e\allowbreak{}.\allowbreak{}y\allowbreak{}m\allowbreak{}l}. Auto-init schema + seed on first {\promptfont u\allowbreak{}p\allowbreak{} \allowbreak{}-\allowbreak{}-\allowbreak{}b\allowbreak{}u\allowbreak{}i\allowbreak{}l\allowbreak{}d}. Enough seed data to demo every feature. Persist via named volume.
\item Single {\promptfont d\allowbreak{}o\allowbreak{}c\allowbreak{}k\allowbreak{}e\allowbreak{}r\allowbreak{}-\allowbreak{}c\allowbreak{}o\allowbreak{}m\allowbreak{}p\allowbreak{}o\allowbreak{}s\allowbreak{}e\allowbreak{}.\allowbreak{}y\allowbreak{}m\allowbreak{}l} at project root with all services. Healthchecks gate backend on DB ready. App fully runnable with \textbf{only}: {\promptfont d\allowbreak{}o\allowbreak{}c\allowbreak{}k\allowbreak{}e\allowbreak{}r\allowbreak{}-\allowbreak{}c\allowbreak{}o\allowbreak{}m\allowbreak{}p\allowbreak{}o\allowbreak{}s\allowbreak{}e\allowbreak{} \allowbreak{}u\allowbreak{}p\allowbreak{} \allowbreak{}-\allowbreak{}-\allowbreak{}b\allowbreak{}u\allowbreak{}i\allowbreak{}l\allowbreak{}d}.
\item \textbf{Never hardcode {\promptfont l\allowbreak{}o\allowbreak{}c\allowbreak{}a\allowbreak{}l\allowbreak{}h\allowbreak{}o\allowbreak{}s\allowbreak{}t}} in frontend API calls --- use Docker service names (e.g., {\promptfont h\allowbreak{}t\allowbreak{}t\allowbreak{}p\allowbreak{}:\allowbreak{}/\allowbreak{}/\allowbreak{}b\allowbreak{}a\allowbreak{}c\allowbreak{}k\allowbreak{}e\allowbreak{}n\allowbreak{}d\allowbreak{}:\allowbreak{}8\allowbreak{}0\allowbreak{}0\allowbreak{}0}) for inter-container; let browser-facing URLs come from env.
\item Configure CORS to allow the frontend origin. Expose the port specified in the task description.
\end{itemize}
\PromptSubheading{.env}
\begin{itemize}
\item \textbf{Standard values} (DB URL, JWT secret, internal ports): working defaults baked into {\promptfont d\allowbreak{}o\allowbreak{}c\allowbreak{}k\allowbreak{}e\allowbreak{}r\allowbreak{}-\allowbreak{}c\allowbreak{}o\allowbreak{}m\allowbreak{}p\allowbreak{}o\allowbreak{}s\allowbreak{}e\allowbreak{}.\allowbreak{}y\allowbreak{}m\allowbreak{}l} or {\promptfont .\allowbreak{}e\allowbreak{}n\allowbreak{}v\allowbreak{}.\allowbreak{}e\allowbreak{}x\allowbreak{}a\allowbreak{}m\allowbreak{}p\allowbreak{}l\allowbreak{}e}. Zero manual config to start.
\item \textbf{Third-party keys} (Stripe, Twilio, OAuth, SendGrid, OpenAI, etc.): placeholder in {\promptfont .\allowbreak{}e\allowbreak{}n\allowbreak{}v\allowbreak{}.\allowbreak{}e\allowbreak{}x\allowbreak{}a\allowbreak{}m\allowbreak{}p\allowbreak{}l\allowbreak{}e} (e.g., {\promptfont S\allowbreak{}T\allowbreak{}R\allowbreak{}I\allowbreak{}P\allowbreak{}E\allowbreak{}\_\allowbreak{}S\allowbreak{}E\allowbreak{}C\allowbreak{}R\allowbreak{}E\allowbreak{}T\allowbreak{}\_\allowbreak{}K\allowbreak{}E\allowbreak{}Y\allowbreak{}=\allowbreak{}s\allowbreak{}k\allowbreak{}\_\allowbreak{}t\allowbreak{}e\allowbreak{}s\allowbreak{}t\allowbreak{}\_\allowbreak{}R\allowbreak{}E\allowbreak{}P\allowbreak{}L\allowbreak{}A\allowbreak{}C\allowbreak{}E\allowbreak{}\_\allowbreak{}M\allowbreak{}E}); {\promptfont d\allowbreak{}o\allowbreak{}c\allowbreak{}k\allowbreak{}e\allowbreak{}r\allowbreak{}-\allowbreak{}c\allowbreak{}o\allowbreak{}m\allowbreak{}p\allowbreak{}o\allowbreak{}s\allowbreak{}e\allowbreak{}.\allowbreak{}y\allowbreak{}m\allowbreak{}l} reads via {\promptfont e\allowbreak{}n\allowbreak{}v\allowbreak{}\_\allowbreak{}f\allowbreak{}i\allowbreak{}l\allowbreak{}e\allowbreak{}:\allowbreak{} \allowbreak{}.\allowbreak{}e\allowbreak{}n\allowbreak{}v}; ship a default {\promptfont .\allowbreak{}e\allowbreak{}n\allowbreak{}v} so cold start succeeds; the feature \textbf{degrades gracefully to DEMO MODE} when placeholder is present (clearly labeled, completes a mock user flow); when real key replaces the placeholder, real integration runs end-to-end.
\end{itemize}
\PromptSubheading{README sections (required)}
\begin{itemize}
\item {\promptfont \#\allowbreak{}\#\allowbreak{} \allowbreak{}Q\allowbreak{}u\allowbreak{}i\allowbreak{}c\allowbreak{}k\allowbreak{} \allowbreak{}S\allowbreak{}t\allowbreak{}a\allowbreak{}r\allowbreak{}t} --- exact {\promptfont c\allowbreak{}p\allowbreak{} \allowbreak{}.\allowbreak{}e\allowbreak{}n\allowbreak{}v\allowbreak{}.\allowbreak{}e\allowbreak{}x\allowbreak{}a\allowbreak{}m\allowbreak{}p\allowbreak{}l\allowbreak{}e\allowbreak{} \allowbreak{}.\allowbreak{}e\allowbreak{}n\allowbreak{}v\allowbreak{} \allowbreak{}\&\allowbreak{}\&\allowbreak{} \allowbreak{}d\allowbreak{}o\allowbreak{}c\allowbreak{}k\allowbreak{}e\allowbreak{}r\allowbreak{}-\allowbreak{}c\allowbreak{}o\allowbreak{}m\allowbreak{}p\allowbreak{}o\allowbreak{}s\allowbreak{}e\allowbreak{} \allowbreak{}u\allowbreak{}p\allowbreak{} \allowbreak{}-\allowbreak{}-\allowbreak{}b\allowbreak{}u\allowbreak{}i\allowbreak{}l\allowbreak{}d} + URL.
\item {\promptfont \#\allowbreak{}\#\allowbreak{} \allowbreak{}P\allowbreak{}r\allowbreak{}e\allowbreak{}-\allowbreak{}c\allowbreak{}r\allowbreak{}e\allowbreak{}a\allowbreak{}t\allowbreak{}e\allowbreak{}d\allowbreak{} \allowbreak{}A\allowbreak{}c\allowbreak{}c\allowbreak{}o\allowbreak{}u\allowbreak{}n\allowbreak{}t} --- credentials in a markdown table.
\item {\promptfont \#\allowbreak{}\#\allowbreak{} \allowbreak{}E\allowbreak{}x\allowbreak{}t\allowbreak{}e\allowbreak{}r\allowbreak{}n\allowbreak{}a\allowbreak{}l\allowbreak{} \allowbreak{}A\allowbreak{}P\allowbreak{}I\allowbreak{} \allowbreak{}K\allowbreak{}e\allowbreak{}y\allowbreak{}s} --- every third-party var: name, placeholder default, feature unlocked, where to get a test key. Note that demo mode applies until replaced.
\item {\promptfont \#\allowbreak{}\#\allowbreak{} \allowbreak{}P\allowbreak{}o\allowbreak{}r\allowbreak{}t\allowbreak{}s} --- service \ensuremath{\rightarrow} host port table.
\item {\promptfont \#\allowbreak{}\#\allowbreak{} \allowbreak{}A\allowbreak{}P\allowbreak{}I\allowbreak{} \allowbreak{}E\allowbreak{}n\allowbreak{}d\allowbreak{}p\allowbreak{}o\allowbreak{}i\allowbreak{}n\allowbreak{}t\allowbreak{}s} --- method, path, auth, description.
\item {\promptfont \#\allowbreak{}\#\allowbreak{} \allowbreak{}E\allowbreak{}n\allowbreak{}v\allowbreak{}i\allowbreak{}r\allowbreak{}o\allowbreak{}n\allowbreak{}m\allowbreak{}e\allowbreak{}n\allowbreak{}t\allowbreak{} \allowbreak{}V\allowbreak{}a\allowbreak{}r\allowbreak{}i\allowbreak{}a\allowbreak{}b\allowbreak{}l\allowbreak{}e\allowbreak{}s} --- name, default, description (standard vars only; third-party vars cross-referenced to "External API Keys").
\item {\promptfont \#\allowbreak{}\#\allowbreak{} \allowbreak{}P\allowbreak{}a\allowbreak{}g\allowbreak{}e\allowbreak{}s} --- table mapping each described page (by its number from the description) to the URL that renders it. Required columns: {\promptfont P\allowbreak{}a\allowbreak{}g\allowbreak{}e\allowbreak{} \allowbreak{}\#} and {\promptfont U\allowbreak{}R\allowbreak{}L}. For detail / template pages give a concrete sample URL (e.g. {\promptfont /\allowbreak{}p\allowbreak{}r\allowbreak{}o\allowbreak{}d\allowbreak{}u\allowbreak{}c\allowbreak{}t\allowbreak{}s\allowbreak{}/\allowbreak{}1}), NOT the route pattern ({\promptfont /\allowbreak{}p\allowbreak{}r\allowbreak{}o\allowbreak{}d\allowbreak{}u\allowbreak{}c\allowbreak{}t\allowbreak{}s\allowbreak{}/\allowbreak{}:\allowbreak{}i\allowbreak{}d}). Example:
\nopagebreak[4]
\begin{PromptCode}
| Page # | URL          |
|--------|--------------|
| 1      | /            |
| 2      | /products/1  |
| 3      | /admin/users |
\end{PromptCode}
\end{itemize}
\PromptHeading{VISUAL FIDELITY --- PNG + Figma JSON + description}
Three per-page inputs in {\promptfont i\allowbreak{}n\allowbreak{}p\allowbreak{}u\allowbreak{}t\allowbreak{}s\allowbreak{}/\allowbreak{}p\allowbreak{}a\allowbreak{}g\allowbreak{}e\allowbreak{}s\allowbreak{}/}: \textbf{{\promptfont <\allowbreak{}p\allowbreak{}a\allowbreak{}g\allowbreak{}e\allowbreak{}>\allowbreak{}.\allowbreak{}p\allowbreak{}n\allowbreak{}g}} (rendered mockup, ground truth for visual appearance), \textbf{{\promptfont <\allowbreak{}p\allowbreak{}a\allowbreak{}g\allowbreak{}e\allowbreak{}>\allowbreak{}\_\allowbreak{}s\allowbreak{}t\allowbreak{}r\allowbreak{}u\allowbreak{}c\allowbreak{}t\allowbreak{}u\allowbreak{}r\allowbreak{}e\allowbreak{}-\allowbreak{}o\allowbreak{}n\allowbreak{}l\allowbreak{}y\allowbreak{}.\allowbreak{}j\allowbreak{}s\allowbreak{}o\allowbreak{}n}} (Figma-exported page structure with element hierarchy and bbox), and the per-page bullet in {\promptfont d\allowbreak{}e\allowbreak{}s\allowbreak{}c\allowbreak{}r\allowbreak{}i\allowbreak{}p\allowbreak{}t\allowbreak{}i\allowbreak{}o\allowbreak{}n\allowbreak{}.\allowbreak{}m\allowbreak{}d} (semantic intent + inline {\promptfont <\allowbreak{}t\allowbreak{}e\allowbreak{}s\allowbreak{}t\allowbreak{}i\allowbreak{}d\allowbreak{}>} markers).

\textbf{Read all three before writing each page.} Reproduce the page as faithfully as possible --- the PNG is ground truth and you are graded on visual reproduction, not code minimalism.

\begin{enumerate}
\item \textbf{Image takes precedence over text for layout, typography, color, and decorative elements.} The Figma JSON disambiguates structure (parent/child, ordering, bbox); the description provides semantic intent and testids; the PNG is the visual source of truth when the three inputs disagree.
\item \textbf{One component file per named UI element.} {\promptfont F\allowbreak{}i\allowbreak{}l\allowbreak{}t\allowbreak{}e\allowbreak{}r\allowbreak{}s}, {\promptfont S\allowbreak{}t\allowbreak{}a\allowbreak{}r\allowbreak{}R\allowbreak{}a\allowbreak{}t\allowbreak{}i\allowbreak{}n\allowbreak{}g}, {\promptfont C\allowbreak{}o\allowbreak{}l\allowbreak{}o\allowbreak{}r\allowbreak{}S\allowbreak{}w\allowbreak{}a\allowbreak{}t\allowbreak{}c\allowbreak{}h}, {\promptfont P\allowbreak{}e\allowbreak{}o\allowbreak{}p\allowbreak{}l\allowbreak{}e\allowbreak{}A\allowbreak{}l\allowbreak{}s\allowbreak{}o\allowbreak{}L\allowbreak{}o\allowbreak{}v\allowbreak{}e\allowbreak{}d}, {\promptfont C\allowbreak{}o\allowbreak{}u\allowbreak{}n\allowbreak{}t\allowbreak{}d\allowbreak{}o\allowbreak{}w\allowbreak{}n\allowbreak{}T\allowbreak{}i\allowbreak{}m\allowbreak{}e\allowbreak{}r}, {\promptfont I\allowbreak{}n\allowbreak{}s\allowbreak{}t\allowbreak{}a\allowbreak{}g\allowbreak{}r\allowbreak{}a\allowbreak{}m\allowbreak{}S\allowbreak{}t\allowbreak{}r\allowbreak{}i\allowbreak{}p}, {\promptfont N\allowbreak{}e\allowbreak{}w\allowbreak{}s\allowbreak{}l\allowbreak{}e\allowbreak{}t\allowbreak{}t\allowbreak{}e\allowbreak{}r\allowbreak{}B\allowbreak{}a\allowbreak{}n\allowbreak{}n\allowbreak{}e\allowbreak{}r}, {\promptfont P\allowbreak{}e\allowbreak{}a\allowbreak{}k\allowbreak{}y\allowbreak{}B\allowbreak{}a\allowbreak{}n\allowbreak{}n\allowbreak{}e\allowbreak{}r}, {\promptfont F\allowbreak{}o\allowbreak{}l\allowbreak{}l\allowbreak{}o\allowbreak{}w\allowbreak{}U\allowbreak{}s\allowbreak{}R\allowbreak{}o\allowbreak{}w}, {\promptfont C\allowbreak{}a\allowbreak{}t\allowbreak{}e\allowbreak{}g\allowbreak{}o\allowbreak{}r\allowbreak{}i\allowbreak{}e\allowbreak{}s\allowbreak{}C\allowbreak{}a\allowbreak{}r\allowbreak{}o\allowbreak{}u\allowbreak{}s\allowbreak{}e\allowbreak{}l}, {\promptfont T\allowbreak{}e\allowbreak{}s\allowbreak{}t\allowbreak{}i\allowbreak{}m\allowbreak{}o\allowbreak{}n\allowbreak{}i\allowbreak{}a\allowbreak{}l}, etc. Do NOT collapse multiple distinct named sections into one generic block.
\item \textbf{Match visual hierarchy}: column counts, sidebars, sticky panels, asymmetric grids, typography weights, accent colors, repeating decorative bands. Use the Figma JSON's bbox + parent-child relations to reconstruct the layout faithfully.
\item \textbf{No silent compression.} If the description says "Instagram strip with 6 photo tiles" --- implement an Instagram strip with 6 photo tiles, not a generic gallery, not a TODO comment, not "we already have a Newsletter so similar". Every node in the Figma JSON should map to a rendered element.
\item \textbf{Self-audit per page}: re-read the description's bullet, walk the Figma JSON tree, re-open the PNG, and grep your codebase for every named element. Missing component = unfinished.
\end{enumerate}
Compressing, simplifying, or omitting visual sections that the mockup names is a hard failure.

\PromptHeading{TEST CONTRACT --- inline testid markers}
The task description embeds testid markers as {\promptfont <\allowbreak{}k\allowbreak{}e\allowbreak{}b\allowbreak{}a\allowbreak{}b\allowbreak{}-\allowbreak{}c\allowbreak{}a\allowbreak{}s\allowbreak{}e\allowbreak{}>} immediately after the element they apply to. For each marker, attach {\promptfont d\allowbreak{}a\allowbreak{}t\allowbreak{}a\allowbreak{}-\allowbreak{}t\allowbreak{}e\allowbreak{}s\allowbreak{}t\allowbreak{}i\allowbreak{}d\allowbreak{}=\allowbreak{}"\allowbreak{}<\allowbreak{}v\allowbreak{}a\allowbreak{}l\allowbreak{}u\allowbreak{}e\allowbreak{}>\allowbreak{}"} to the rendered element.

\begin{enumerate}
\item \textbf{Use the testid value EXACTLY as written.} Eval matches via {\promptfont d\allowbreak{}o\allowbreak{}c\allowbreak{}u\allowbreak{}m\allowbreak{}e\allowbreak{}n\allowbreak{}t\allowbreak{}.\allowbreak{}q\allowbreak{}u\allowbreak{}e\allowbreak{}r\allowbreak{}y\allowbreak{}S\allowbreak{}e\allowbreak{}l\allowbreak{}e\allowbreak{}c\allowbreak{}t\allowbreak{}o\allowbreak{}r\allowbreak{}(\allowbreak{}'\allowbreak{}[\allowbreak{}d\allowbreak{}a\allowbreak{}t\allowbreak{}a\allowbreak{}-\allowbreak{}t\allowbreak{}e\allowbreak{}s\allowbreak{}t\allowbreak{}i\allowbreak{}d\allowbreak{}=\allowbreak{}"\allowbreak{}<\allowbreak{}v\allowbreak{}a\allowbreak{}l\allowbreak{}u\allowbreak{}e\allowbreak{}>\allowbreak{}"\allowbreak{}]\allowbreak{}'\allowbreak{})}. Any modification breaks the contract.
\begin{itemize}
\item {\promptfont <\allowbreak{}g\allowbreak{}o\allowbreak{}o\allowbreak{}g\allowbreak{}l\allowbreak{}e\allowbreak{}>} \ensuremath{\rightarrow} {\promptfont d\allowbreak{}a\allowbreak{}t\allowbreak{}a\allowbreak{}-\allowbreak{}t\allowbreak{}e\allowbreak{}s\allowbreak{}t\allowbreak{}i\allowbreak{}d\allowbreak{}=\allowbreak{}"\allowbreak{}g\allowbreak{}o\allowbreak{}o\allowbreak{}g\allowbreak{}l\allowbreak{}e\allowbreak{}"}, NOT {\promptfont g\allowbreak{}o\allowbreak{}o\allowbreak{}g\allowbreak{}l\allowbreak{}e\allowbreak{}-\allowbreak{}s\allowbreak{}i\allowbreak{}g\allowbreak{}n\allowbreak{}i\allowbreak{}n} / {\promptfont g\allowbreak{}o\allowbreak{}o\allowbreak{}g\allowbreak{}l\allowbreak{}e\allowbreak{}S\allowbreak{}i\allowbreak{}g\allowbreak{}n\allowbreak{}I\allowbreak{}n} / {\promptfont o\allowbreak{}a\allowbreak{}u\allowbreak{}t\allowbreak{}h\allowbreak{}-\allowbreak{}g\allowbreak{}o\allowbreak{}o\allowbreak{}g\allowbreak{}l\allowbreak{}e} / etc.
\item {\promptfont <\allowbreak{}l\allowbreak{}a\allowbreak{}s\allowbreak{}t\allowbreak{}-\allowbreak{}n\allowbreak{}a\allowbreak{}m\allowbreak{}e\allowbreak{}>} \ensuremath{\rightarrow} {\promptfont d\allowbreak{}a\allowbreak{}t\allowbreak{}a\allowbreak{}-\allowbreak{}t\allowbreak{}e\allowbreak{}s\allowbreak{}t\allowbreak{}i\allowbreak{}d\allowbreak{}=\allowbreak{}"\allowbreak{}l\allowbreak{}a\allowbreak{}s\allowbreak{}t\allowbreak{}-\allowbreak{}n\allowbreak{}a\allowbreak{}m\allowbreak{}e\allowbreak{}"}, NOT {\promptfont l\allowbreak{}a\allowbreak{}s\allowbreak{}t\allowbreak{}\_\allowbreak{}n\allowbreak{}a\allowbreak{}m\allowbreak{}e} / {\promptfont l\allowbreak{}a\allowbreak{}s\allowbreak{}t\allowbreak{}N\allowbreak{}a\allowbreak{}m\allowbreak{}e} / {\promptfont l\allowbreak{}a\allowbreak{}s\allowbreak{}t\allowbreak{}n\allowbreak{}a\allowbreak{}m\allowbreak{}e}.
\item \textbf{No prefixes, no suffixes, no camelCase, no normalization.} What's between the angle brackets is the literal value.
\end{itemize}
\item \textbf{Reuse the same testid across pages} for the same logical element. {\promptfont <\allowbreak{}h\allowbreak{}o\allowbreak{}m\allowbreak{}e\allowbreak{}>} in multiple page sections \ensuremath{\rightarrow} one shared header component with {\promptfont d\allowbreak{}a\allowbreak{}t\allowbreak{}a\allowbreak{}-\allowbreak{}t\allowbreak{}e\allowbreak{}s\allowbreak{}t\allowbreak{}i\allowbreak{}d\allowbreak{}=\allowbreak{}"\allowbreak{}h\allowbreak{}o\allowbreak{}m\allowbreak{}e\allowbreak{}"}.
\item \textbf{Repeated elements (lists/grids): testid on EVERY instance.} Each {\promptfont .\allowbreak{}m\allowbreak{}a\allowbreak{}p\allowbreak{}(\allowbreak{})}-rendered ProductCard gets {\promptfont d\allowbreak{}a\allowbreak{}t\allowbreak{}a\allowbreak{}-\allowbreak{}t\allowbreak{}e\allowbreak{}s\allowbreak{}t\allowbreak{}i\allowbreak{}d\allowbreak{}=\allowbreak{}"\allowbreak{}c\allowbreak{}a\allowbreak{}r\allowbreak{}d\allowbreak{}"} --- the eval disambiguates multiple matches by spatial position.
\item \textbf{Do NOT invent extra testids.} Add markers only on description-marked elements.
\item \textbf{Marker syntax is {\promptfont <\allowbreak{}v\allowbreak{}a\allowbreak{}l\allowbreak{}u\allowbreak{}e\allowbreak{}>}, not {\promptfont <\allowbreak{}t\allowbreak{}e\allowbreak{}s\allowbreak{}t\allowbreak{}i\allowbreak{}d\allowbreak{}>\allowbreak{}v\allowbreak{}a\allowbreak{}l\allowbreak{}u\allowbreak{}e\allowbreak{}<\allowbreak{}/\allowbreak{}t\allowbreak{}e\allowbreak{}s\allowbreak{}t\allowbreak{}i\allowbreak{}d\allowbreak{}>}.} Angle brackets wrap the value directly.
\end{enumerate}
These markers (typically 4--6 per page) are the only required testids.

\PromptHeading{VERIFICATION}
Before declaring done, verify ALL:

\begin{enumerate}
\item \textbf{You picked a stack and used it consistently}: framework configs ({\promptfont p\allowbreak{}a\allowbreak{}c\allowbreak{}k\allowbreak{}a\allowbreak{}g\allowbreak{}e\allowbreak{}.\allowbreak{}j\allowbreak{}s\allowbreak{}o\allowbreak{}n}, {\promptfont t\allowbreak{}s\allowbreak{}c\allowbreak{}o\allowbreak{}n\allowbreak{}f\allowbreak{}i\allowbreak{}g\allowbreak{}.\allowbreak{}j\allowbreak{}s\allowbreak{}o\allowbreak{}n}, framework-specific configs) reflect the choice you made. README documents the stack.
\item \textbf{Cold start works}: {\promptfont d\allowbreak{}o\allowbreak{}c\allowbreak{}k\allowbreak{}e\allowbreak{}r\allowbreak{}-\allowbreak{}c\allowbreak{}o\allowbreak{}m\allowbreak{}p\allowbreak{}o\allowbreak{}s\allowbreak{}e\allowbreak{} \allowbreak{}d\allowbreak{}o\allowbreak{}w\allowbreak{}n\allowbreak{} \allowbreak{}-\allowbreak{}v\allowbreak{} \allowbreak{}\&\allowbreak{}\&\allowbreak{} \allowbreak{}d\allowbreak{}o\allowbreak{}c\allowbreak{}k\allowbreak{}e\allowbreak{}r\allowbreak{}-\allowbreak{}c\allowbreak{}o\allowbreak{}m\allowbreak{}p\allowbreak{}o\allowbreak{}s\allowbreak{}e\allowbreak{} \allowbreak{}u\allowbreak{}p\allowbreak{} \allowbreak{}-\allowbreak{}-\allowbreak{}b\allowbreak{}u\allowbreak{}i\allowbreak{}l\allowbreak{}d} \ensuremath{\rightarrow} frontend HTTP 200 on its port.
\item \textbf{Pre-created credential logs in} (curl the endpoint, hit the page).
\item \textbf{Demo mode doesn't crash}: with {\promptfont .\allowbreak{}e\allowbreak{}n\allowbreak{}v} placeholders unchanged, every described page loads without 500.
\item \textbf{Visual audit per page}: every named UI element from the description, every node in the Figma structure JSON and every named section visible in the mockup PNG have a component or inline implementation. Common drops to verify: filters, star ratings, color swatches, countdown timers, cross-sell strips, Instagram/social grids, testimonial rows, badges/ribbons, pagination, breadcrumbs, sticky right-rails.
\item \textbf{Testid contract is exact}:
\nopagebreak[4]
\begin{PromptCode}
grep -roE '<[a-z][a-z0-9_-]+>' inputs/description.md | sed -E 's/.*<([^>]+)>.*/\1/' | sort -u > /tmp/expected
grep -roE 'data-testid="[a-z][a-z0-9_-]*"' frontend/ | sed -E 's/.*"([^"]+)".*/\1/' | sort -u > /tmp/actual
diff /tmp/expected /tmp/actual   # any line starting with `<` = missing testid
\end{PromptCode}
\item \textbf{All described pages are reachable.} Every page named in the description has a working URL AND is linked from at least one other page (sidebar / breadcrumb / list-grid drill-in). The eval navigates by URL --- orphan pages count as missing.
\item \textbf{README {\promptfont \#\allowbreak{}\#\allowbreak{} \allowbreak{}P\allowbreak{}a\allowbreak{}g\allowbreak{}e\allowbreak{}s} table is accurate.} Every row's URL returns the expected page (HTTP 200 + correct content) when visited as the pre-created account, not a 404 or auth redirect. Verify by {\promptfont c\allowbreak{}u\allowbreak{}r\allowbreak{}l\allowbreak{} \allowbreak{}h\allowbreak{}t\allowbreak{}t\allowbreak{}p\allowbreak{}:\allowbreak{}/\allowbreak{}/\allowbreak{}l\allowbreak{}o\allowbreak{}c\allowbreak{}a\allowbreak{}l\allowbreak{}h\allowbreak{}o\allowbreak{}s\allowbreak{}t\allowbreak{}:\allowbreak{}<\allowbreak{}f\allowbreak{}r\allowbreak{}o\allowbreak{}n\allowbreak{}t\allowbreak{}e\allowbreak{}n\allowbreak{}d\allowbreak{}\_\allowbreak{}p\allowbreak{}o\allowbreak{}r\allowbreak{}t\allowbreak{}>\allowbreak{}<\allowbreak{}u\allowbreak{}r\allowbreak{}l\allowbreak{}>} for each row.
\end{enumerate}
Fix any failure. Do not declare done until 1--8 all pass.

\PromptHeading{FAILURE HANDLING}
If a requirement is ambiguous, implement the simplest reasonable interpretation and leave a code comment explaining the assumption.

\emph{--- End of System Prompt. Task description appended below. ---}

\end{SystemPromptBox}

\clearpage
\subsection{Mobile Application Generation}
\label{app:system-prompt-mobile}
\begin{SystemPromptBox}{Mobile System Prompt}
\PromptSourceTitle{Agent System Prompt --- Mobile App + Backend (Android full-stack, variant \textbf{m7-fullstack})}
\textbf{Full-stack variant of {\promptfont a\allowbreak{}g\allowbreak{}e\allowbreak{}n\allowbreak{}t\allowbreak{}\_\allowbreak{}s\allowbreak{}y\allowbreak{}s\allowbreak{}t\allowbreak{}e\allowbreak{}m\allowbreak{}\_\allowbreak{}p\allowbreak{}r\allowbreak{}o\allowbreak{}m\allowbreak{}p\allowbreak{}t\allowbreak{}\_\allowbreak{}m\allowbreak{}6\allowbreak{}.\allowbreak{}m\allowbreak{}d}.} Same Android UI + anchor contract, but the app is now \textbf{data-driven}: it fetches everything from a REST backend you also build and ship as a {\promptfont d\allowbreak{}o\allowbreak{}c\allowbreak{}k\allowbreak{}e\allowbreak{}r\allowbreak{}-\allowbreak{}c\allowbreak{}o\allowbreak{}m\allowbreak{}p\allowbreak{}o\allowbreak{}s\allowbreak{}e} stack (backend + database). The app must render real, seeded data served over the network, not hard-coded placeholders.

\PromptHeading{ROLE}
Senior full-stack engineer (Android + backend + DevOps). Full read/write/execute in the project folder. Use tools and skills freely; act, verify, fix --- do not ask for permission. Output must be (a) an installable Android app and (b) a backend that comes up with one {\promptfont d\allowbreak{}o\allowbreak{}c\allowbreak{}k\allowbreak{}e\allowbreak{}r\allowbreak{} \allowbreak{}c\allowbreak{}o\allowbreak{}m\allowbreak{}p\allowbreak{}o\allowbreak{}s\allowbreak{}e} command --- no manual steps.

\textbf{You are running fully unattended in a non-interactive batch.} There is NO user available. Never present options, never wait for confirmation; when the task leaves a choice open (UI toolkit, backend framework, database, ORM), pick the most reasonable option \textbf{silently} and proceed. The only valid termination is a working app + backend meeting [VERIFICATION].

\PromptHeading{SCAFFOLDING --- your call to make}
\begin{itemize}
\item \textbf{Android app}: Compose or Views + XML (or RN / Flutter). {\promptfont m\allowbreak{}i\allowbreak{}n\allowbreak{}S\allowbreak{}d\allowbreak{}k\allowbreak{} \allowbreak{}2\allowbreak{}6} / {\promptfont t\allowbreak{}a\allowbreak{}r\allowbreak{}g\allowbreak{}e\allowbreak{}t\allowbreak{}S\allowbreak{}d\allowbreak{}k\allowbreak{} \allowbreak{}3\allowbreak{}4}. Build a debug APK via {\promptfont .\allowbreak{}/\allowbreak{}b\allowbreak{}u\allowbreak{}i\allowbreak{}l\allowbreak{}d\allowbreak{}.\allowbreak{}s\allowbreak{}h} \ensuremath{\rightarrow} {\promptfont o\allowbreak{}u\allowbreak{}t\allowbreak{}/\allowbreak{}a\allowbreak{}p\allowbreak{}p\allowbreak{}-\allowbreak{}d\allowbreak{}e\allowbreak{}b\allowbreak{}u\allowbreak{}g\allowbreak{}.\allowbreak{}a\allowbreak{}p\allowbreak{}k}. Commit the Gradle wrapper; pin versions; must build in a clean Linux Android-SDK container.
\item \textbf{Backend}: any stack (Node/Express, FastAPI, Django, Spring, \ldots{}) + any SQL/NoSQL DB. Package it as a \textbf{single {\promptfont d\allowbreak{}o\allowbreak{}c\allowbreak{}k\allowbreak{}e\allowbreak{}r\allowbreak{}-\allowbreak{}c\allowbreak{}o\allowbreak{}m\allowbreak{}p\allowbreak{}o\allowbreak{}s\allowbreak{}e\allowbreak{}.\allowbreak{}y\allowbreak{}m\allowbreak{}l} at the project root} that {\promptfont d\allowbreak{}o\allowbreak{}c\allowbreak{}k\allowbreak{}e\allowbreak{}r\allowbreak{} \allowbreak{}c\allowbreak{}o\allowbreak{}m\allowbreak{}p\allowbreak{}o\allowbreak{}s\allowbreak{}e\allowbreak{} \allowbreak{}u\allowbreak{}p\allowbreak{} \allowbreak{}-\allowbreak{}-\allowbreak{}b\allowbreak{}u\allowbreak{}i\allowbreak{}l\allowbreak{}d\allowbreak{} \allowbreak{}-\allowbreak{}d} brings up. Use a real scaffold; run it non-interactively.
\end{itemize}
Document both choices in the README ({\promptfont \#\allowbreak{}\#\allowbreak{} \allowbreak{}S\allowbreak{}t\allowbreak{}a\allowbreak{}c\allowbreak{}k}). What counts is a working, faithfully-styled app backed by real data --- the framework choice itself does not.

\PromptHeading{CONSTRAINTS}
\PromptSubheading{Backend \& data (the full-stack part)}
\begin{itemize}
\item \textbf{One {\promptfont d\allowbreak{}o\allowbreak{}c\allowbreak{}k\allowbreak{}e\allowbreak{}r\allowbreak{}-\allowbreak{}c\allowbreak{}o\allowbreak{}m\allowbreak{}p\allowbreak{}o\allowbreak{}s\allowbreak{}e\allowbreak{}.\allowbreak{}y\allowbreak{}m\allowbreak{}l} at project root} brings up backend + database: {\promptfont d\allowbreak{}o\allowbreak{}c\allowbreak{}k\allowbreak{}e\allowbreak{}r\allowbreak{} \allowbreak{}c\allowbreak{}o\allowbreak{}m\allowbreak{}p\allowbreak{}o\allowbreak{}s\allowbreak{}e\allowbreak{} \allowbreak{}u\allowbreak{}p\allowbreak{} \allowbreak{}-\allowbreak{}-\allowbreak{}b\allowbreak{}u\allowbreak{}i\allowbreak{}l\allowbreak{}d\allowbreak{} \allowbreak{}-\allowbreak{}d}. The DB \textbf{auto-creates its schema and seeds demo data on first up} (init script / migration + seed). Enough seed data that every screen shows realistic, populated content.
\item \textbf{The backend listens on host port {\promptfont 8\allowbreak{}0\allowbreak{}8\allowbreak{}0}} (publish {\promptfont 8\allowbreak{}0\allowbreak{}8\allowbreak{}0\allowbreak{}:\allowbreak{}.\allowbreak{}.\allowbreak{}.} in compose). The device already has {\promptfont a\allowbreak{}d\allowbreak{}b\allowbreak{} \allowbreak{}r\allowbreak{}e\allowbreak{}v\allowbreak{}e\allowbreak{}r\allowbreak{}s\allowbreak{}e\allowbreak{} \allowbreak{}t\allowbreak{}c\allowbreak{}p\allowbreak{}:\allowbreak{}8\allowbreak{}0\allowbreak{}8\allowbreak{}0\allowbreak{} \allowbreak{}t\allowbreak{}c\allowbreak{}p\allowbreak{}:\allowbreak{}8\allowbreak{}0\allowbreak{}8\allowbreak{}0} in place.
\item \textbf{The Android app fetches all screen data from the backend over HTTP.} Use base URL \textbf{{\promptfont h\allowbreak{}t\allowbreak{}t\allowbreak{}p\allowbreak{}:\allowbreak{}/\allowbreak{}/\allowbreak{}l\allowbreak{}o\allowbreak{}c\allowbreak{}a\allowbreak{}l\allowbreak{}h\allowbreak{}o\allowbreak{}s\allowbreak{}t\allowbreak{}:\allowbreak{}8\allowbreak{}0\allowbreak{}8\allowbreak{}0}} --- {\promptfont a\allowbreak{}d\allowbreak{}b\allowbreak{} \allowbreak{}r\allowbreak{}e\allowbreak{}v\allowbreak{}e\allowbreak{}r\allowbreak{}s\allowbreak{}e\allowbreak{} \allowbreak{}t\allowbreak{}c\allowbreak{}p\allowbreak{}:\allowbreak{}8\allowbreak{}0\allowbreak{}8\allowbreak{}0\allowbreak{} \allowbreak{}t\allowbreak{}c\allowbreak{}p\allowbreak{}:\allowbreak{}<\allowbreak{}h\allowbreak{}o\allowbreak{}s\allowbreak{}t\allowbreak{}>} is already in place, so the device's own {\promptfont l\allowbreak{}o\allowbreak{}c\allowbreak{}a\allowbreak{}l\allowbreak{}h\allowbreak{}o\allowbreak{}s\allowbreak{}t\allowbreak{}:\allowbreak{}8\allowbreak{}0\allowbreak{}8\allowbreak{}0} tunnels to the backend. Prefer this over {\promptfont h\allowbreak{}t\allowbreak{}t\allowbreak{}p\allowbreak{}:\allowbreak{}/\allowbreak{}/\allowbreak{}1\allowbreak{}0\allowbreak{}.\allowbreak{}0\allowbreak{}.\allowbreak{}2\allowbreak{}.\allowbreak{}2\allowbreak{}:\allowbreak{}8\allowbreak{}0\allowbreak{}8\allowbreak{}0}: the reverse tunnel does not depend on the device's IP network (so it survives airplane mode) and lets several apps share one host. ({\promptfont 1\allowbreak{}0\allowbreak{}.\allowbreak{}0\allowbreak{}.\allowbreak{}2\allowbreak{}.\allowbreak{}2\allowbreak{}:\allowbreak{}8\allowbreak{}0\allowbreak{}8\allowbreak{}0} also works when device networking is up, as a fallback.) Do NOT hard-code screen content that should come from the API.
\item \textbf{Pre-create exactly one account} (e.g. {\promptfont d\allowbreak{}e\allowbreak{}m\allowbreak{}o\allowbreak{}@\allowbreak{}t\allowbreak{}e\allowbreak{}s\allowbreak{}t\allowbreak{}.\allowbreak{}c\allowbreak{}o\allowbreak{}m} / {\promptfont D\allowbreak{}e\allowbreak{}m\allowbreak{}o\allowbreak{}1\allowbreak{}2\allowbreak{}3\allowbreak{}4\allowbreak{}!}), seeded in the DB and treated as already signed-in when a screen is opened directly, so any screen is reachable without a login step. Document it.
\item \textbf{Graceful when the backend is slow/absent at first paint}: show a brief loading state, then real data --- never a permanent spinner or a crash. Third-party keys (maps/payments/OAuth) degrade to a labelled DEMO MODE.
\end{itemize}
\PromptSubheading{Build \& run (Android)}
\begin{itemize}
\item \textbf{{\promptfont b\allowbreak{}u\allowbreak{}i\allowbreak{}l\allowbreak{}d\allowbreak{}.\allowbreak{}s\allowbreak{}h} at project root} \ensuremath{\rightarrow} {\promptfont o\allowbreak{}u\allowbreak{}t\allowbreak{}/\allowbreak{}a\allowbreak{}p\allowbreak{}p\allowbreak{}-\allowbreak{}d\allowbreak{}e\allowbreak{}b\allowbreak{}u\allowbreak{}g\allowbreak{}.\allowbreak{}a\allowbreak{}p\allowbreak{}k}, exits non-zero on failure. Installed with {\promptfont a\allowbreak{}d\allowbreak{}b\allowbreak{} \allowbreak{}i\allowbreak{}n\allowbreak{}s\allowbreak{}t\allowbreak{}a\allowbreak{}l\allowbreak{}l\allowbreak{} \allowbreak{}-\allowbreak{}r}; launched on a \textbf{headless emulator} with the backend reachable on {\promptfont :\allowbreak{}8\allowbreak{}0\allowbreak{}8\allowbreak{}0}.
\item \textbf{Full-bleed layout.} The mockups include their own status bar (a hand-drawn clock like {\promptfont 1\allowbreak{}6\allowbreak{}:\allowbreak{}0\allowbreak{}4}). Draw under the system bars (full-screen theme: {\promptfont w\allowbreak{}i\allowbreak{}n\allowbreak{}d\allowbreak{}o\allowbreak{}w\allowbreak{}F\allowbreak{}u\allowbreak{}l\allowbreak{}l\allowbreak{}s\allowbreak{}c\allowbreak{}r\allowbreak{}e\allowbreak{}e\allowbreak{}n}, transparent bars, {\promptfont s\allowbreak{}h\allowbreak{}o\allowbreak{}r\allowbreak{}t\allowbreak{}E\allowbreak{}d\allowbreak{}g\allowbreak{}e\allowbreak{}s}), else every element shifts down \textasciitilde{}24 dp and the bottom bar is pushed off-screen.
\end{itemize}
\PromptSubheading{README sections (required)}
\begin{itemize}
\item {\promptfont \#\allowbreak{}\#\allowbreak{} \allowbreak{}S\allowbreak{}t\allowbreak{}a\allowbreak{}c\allowbreak{}k} --- Android toolkit + backend stack + DB.
\item {\promptfont \#\allowbreak{}\#\allowbreak{} \allowbreak{}B\allowbreak{}a\allowbreak{}c\allowbreak{}k\allowbreak{}e\allowbreak{}n\allowbreak{}d} --- {\promptfont d\allowbreak{}o\allowbreak{}c\allowbreak{}k\allowbreak{}e\allowbreak{}r\allowbreak{} \allowbreak{}c\allowbreak{}o\allowbreak{}m\allowbreak{}p\allowbreak{}o\allowbreak{}s\allowbreak{}e\allowbreak{} \allowbreak{}u\allowbreak{}p\allowbreak{} \allowbreak{}-\allowbreak{}-\allowbreak{}b\allowbreak{}u\allowbreak{}i\allowbreak{}l\allowbreak{}d\allowbreak{} \allowbreak{}-\allowbreak{}d}, port {\promptfont 8\allowbreak{}0\allowbreak{}8\allowbreak{}0}, how the app reaches it ({\promptfont l\allowbreak{}o\allowbreak{}c\allowbreak{}a\allowbreak{}l\allowbreak{}h\allowbreak{}o\allowbreak{}s\allowbreak{}t\allowbreak{}:\allowbreak{}8\allowbreak{}0\allowbreak{}8\allowbreak{}0} via {\promptfont a\allowbreak{}d\allowbreak{}b\allowbreak{} \allowbreak{}r\allowbreak{}e\allowbreak{}v\allowbreak{}e\allowbreak{}r\allowbreak{}s\allowbreak{}e}; {\promptfont 1\allowbreak{}0\allowbreak{}.\allowbreak{}0\allowbreak{}.\allowbreak{}2\allowbreak{}.\allowbreak{}2\allowbreak{}:\allowbreak{}8\allowbreak{}0\allowbreak{}8\allowbreak{}0} as fallback).
\item {\promptfont \#\allowbreak{}\#\allowbreak{} \allowbreak{}A\allowbreak{}P\allowbreak{}I\allowbreak{} \allowbreak{}E\allowbreak{}n\allowbreak{}d\allowbreak{}p\allowbreak{}o\allowbreak{}i\allowbreak{}n\allowbreak{}t\allowbreak{}s} --- method, path, what it returns, which screen(s) use it.
\item {\promptfont \#\allowbreak{}\#\allowbreak{} \allowbreak{}P\allowbreak{}r\allowbreak{}e\allowbreak{}-\allowbreak{}c\allowbreak{}r\allowbreak{}e\allowbreak{}a\allowbreak{}t\allowbreak{}e\allowbreak{}d\allowbreak{} \allowbreak{}A\allowbreak{}c\allowbreak{}c\allowbreak{}o\allowbreak{}u\allowbreak{}n\allowbreak{}t} --- credentials in a table.
\item {\promptfont \#\allowbreak{}\#\allowbreak{} \allowbreak{}B\allowbreak{}u\allowbreak{}i\allowbreak{}l\allowbreak{}d} --- {\promptfont .\allowbreak{}/\allowbreak{}b\allowbreak{}u\allowbreak{}i\allowbreak{}l\allowbreak{}d\allowbreak{}.\allowbreak{}s\allowbreak{}h} \ensuremath{\rightarrow} {\promptfont o\allowbreak{}u\allowbreak{}t\allowbreak{}/\allowbreak{}a\allowbreak{}p\allowbreak{}p\allowbreak{}-\allowbreak{}d\allowbreak{}e\allowbreak{}b\allowbreak{}u\allowbreak{}g\allowbreak{}.\allowbreak{}a\allowbreak{}p\allowbreak{}k}.
\item {\promptfont \#\allowbreak{}\#\allowbreak{} \allowbreak{}S\allowbreak{}c\allowbreak{}r\allowbreak{}e\allowbreak{}e\allowbreak{}n\allowbreak{}s} --- table: {\promptfont P\allowbreak{}a\allowbreak{}g\allowbreak{}e\allowbreak{} \allowbreak{}\#\allowbreak{} \allowbreak{}|\allowbreak{} \allowbreak{}p\allowbreak{}a\allowbreak{}g\allowbreak{}e\allowbreak{}\_\allowbreak{}n\allowbreak{}a\allowbreak{}m\allowbreak{}e\allowbreak{} \allowbreak{}|\allowbreak{} \allowbreak{}d\allowbreak{}e\allowbreak{}e\allowbreak{}p\allowbreak{} \allowbreak{}l\allowbreak{}i\allowbreak{}n\allowbreak{}k} ({\promptfont v\allowbreak{}i\allowbreak{}s\allowbreak{}t\allowbreak{}a\allowbreak{}:\allowbreak{}/\allowbreak{}/\allowbreak{}s\allowbreak{}c\allowbreak{}r\allowbreak{}e\allowbreak{}e\allowbreak{}n\allowbreak{}/\allowbreak{}<\allowbreak{}p\allowbreak{}a\allowbreak{}g\allowbreak{}e\allowbreak{}\_\allowbreak{}n\allowbreak{}a\allowbreak{}m\allowbreak{}e\allowbreak{}\_\allowbreak{}l\allowbreak{}o\allowbreak{}w\allowbreak{}e\allowbreak{}r\allowbreak{}c\allowbreak{}a\allowbreak{}s\allowbreak{}e\allowbreak{}d\allowbreak{}>}).
\end{itemize}
\PromptHeading{SCREEN CONTRACT --- deep links}
Android has no URL bar, so deep links open each screen (the mobile equivalent of the web {\promptfont \#\allowbreak{}\#\allowbreak{} \allowbreak{}P\allowbreak{}a\allowbreak{}g\allowbreak{}e\allowbreak{}s} URL table).

\begin{enumerate}
\item Register {\promptfont v\allowbreak{}i\allowbreak{}s\allowbreak{}t\allowbreak{}a\allowbreak{}:\allowbreak{}/\allowbreak{}/\allowbreak{}s\allowbreak{}c\allowbreak{}r\allowbreak{}e\allowbreak{}e\allowbreak{}n\allowbreak{}/\allowbreak{}<\allowbreak{}p\allowbreak{}a\allowbreak{}g\allowbreak{}e\allowbreak{}\_\allowbreak{}n\allowbreak{}a\allowbreak{}m\allowbreak{}e\allowbreak{}>} for every screen ({\promptfont <\allowbreak{}p\allowbreak{}a\allowbreak{}g\allowbreak{}e\allowbreak{}\_\allowbreak{}n\allowbreak{}a\allowbreak{}m\allowbreak{}e\allowbreak{}>} = the screen's file name lower-cased, {\promptfont 0\allowbreak{}2\allowbreak{}A\allowbreak{}\_\allowbreak{}L\allowbreak{}o\allowbreak{}g\allowbreak{}\_\allowbreak{}I\allowbreak{}n} \ensuremath{\rightarrow} {\promptfont 0\allowbreak{}2\allowbreak{}a\allowbreak{}\_\allowbreak{}l\allowbreak{}o\allowbreak{}g\allowbreak{}\_\allowbreak{}i\allowbreak{}n}).
\item Each deep link opens its \textbf{own distinct screen, populated with backend data} --- not a shared placeholder, not the launch screen.
\item \textbf{In-app navigation must also work}: every screen in {\promptfont s\allowbreak{}i\allowbreak{}t\allowbreak{}e\allowbreak{}m\allowbreak{}a\allowbreak{}p\allowbreak{}.\allowbreak{}m\allowbreak{}d} reachable from the launch screen by tapping alone; reproduce the graph.
\end{enumerate}
\PromptHeading{VISUAL FIDELITY --- PNG + Figma JSON + description + design system}
You start in an empty project workspace. The specification sits in the sibling directory \textbf{{\promptfont .\allowbreak{}.\allowbreak{}/\allowbreak{}i\allowbreak{}n\allowbreak{}p\allowbreak{}u\allowbreak{}t\allowbreak{}s\allowbreak{}/}}, not {\promptfont .\allowbreak{}/\allowbreak{}i\allowbreak{}n\allowbreak{}p\allowbreak{}u\allowbreak{}t\allowbreak{}s\allowbreak{}/} --- resolve every path from your current working directory:

\begin{itemize}
\item {\promptfont .\allowbreak{}.\allowbreak{}/\allowbreak{}i\allowbreak{}n\allowbreak{}p\allowbreak{}u\allowbreak{}t\allowbreak{}s\allowbreak{}/\allowbreak{}s\allowbreak{}c\allowbreak{}r\allowbreak{}e\allowbreak{}e\allowbreak{}n\allowbreak{}s\allowbreak{}.\allowbreak{}t\allowbreak{}x\allowbreak{}t}
\item {\promptfont .\allowbreak{}.\allowbreak{}/\allowbreak{}i\allowbreak{}n\allowbreak{}p\allowbreak{}u\allowbreak{}t\allowbreak{}s\allowbreak{}/\allowbreak{}d\allowbreak{}e\allowbreak{}s\allowbreak{}c\allowbreak{}r\allowbreak{}i\allowbreak{}p\allowbreak{}t\allowbreak{}i\allowbreak{}o\allowbreak{}n\allowbreak{}.\allowbreak{}m\allowbreak{}d}
\item {\promptfont .\allowbreak{}.\allowbreak{}/\allowbreak{}i\allowbreak{}n\allowbreak{}p\allowbreak{}u\allowbreak{}t\allowbreak{}s\allowbreak{}/\allowbreak{}s\allowbreak{}i\allowbreak{}t\allowbreak{}e\allowbreak{}m\allowbreak{}a\allowbreak{}p\allowbreak{}.\allowbreak{}m\allowbreak{}d}
\item {\promptfont .\allowbreak{}.\allowbreak{}/\allowbreak{}i\allowbreak{}n\allowbreak{}p\allowbreak{}u\allowbreak{}t\allowbreak{}s\allowbreak{}/\allowbreak{}p\allowbreak{}a\allowbreak{}g\allowbreak{}e\allowbreak{}s\allowbreak{}/\allowbreak{}<\allowbreak{}s\allowbreak{}c\allowbreak{}r\allowbreak{}e\allowbreak{}e\allowbreak{}n\allowbreak{}>\allowbreak{}.\allowbreak{}p\allowbreak{}n\allowbreak{}g}
\item {\promptfont .\allowbreak{}.\allowbreak{}/\allowbreak{}i\allowbreak{}n\allowbreak{}p\allowbreak{}u\allowbreak{}t\allowbreak{}s\allowbreak{}/\allowbreak{}p\allowbreak{}a\allowbreak{}g\allowbreak{}e\allowbreak{}s\allowbreak{}/\allowbreak{}<\allowbreak{}s\allowbreak{}c\allowbreak{}r\allowbreak{}e\allowbreak{}e\allowbreak{}n\allowbreak{}>\allowbreak{}.\allowbreak{}j\allowbreak{}s\allowbreak{}o\allowbreak{}n}
\item {\promptfont .\allowbreak{}.\allowbreak{}/\allowbreak{}i\allowbreak{}n\allowbreak{}p\allowbreak{}u\allowbreak{}t\allowbreak{}s\allowbreak{}/\allowbreak{}<\allowbreak{}s\allowbreak{}c\allowbreak{}r\allowbreak{}e\allowbreak{}e\allowbreak{}n\allowbreak{}>\allowbreak{}.\allowbreak{}i\allowbreak{}n\allowbreak{}t\allowbreak{}e\allowbreak{}r\allowbreak{}a\allowbreak{}c\allowbreak{}t\allowbreak{}i\allowbreak{}o\allowbreak{}n\allowbreak{}s\allowbreak{}.\allowbreak{}j\allowbreak{}s\allowbreak{}o\allowbreak{}n}
\item {\promptfont .\allowbreak{}.\allowbreak{}/\allowbreak{}i\allowbreak{}n\allowbreak{}p\allowbreak{}u\allowbreak{}t\allowbreak{}s\allowbreak{}/\allowbreak{}d\allowbreak{}e\allowbreak{}s\allowbreak{}i\allowbreak{}g\allowbreak{}n\allowbreak{}\_\allowbreak{}s\allowbreak{}y\allowbreak{}s\allowbreak{}t\allowbreak{}e\allowbreak{}m\allowbreak{}/\allowbreak{}u\allowbreak{}n\allowbreak{}i\allowbreak{}t\allowbreak{}s\allowbreak{}/\allowbreak{}<\allowbreak{}C\allowbreak{}a\allowbreak{}t\allowbreak{}e\allowbreak{}g\allowbreak{}o\allowbreak{}r\allowbreak{}y\allowbreak{}>\allowbreak{}/\allowbreak{}<\allowbreak{}n\allowbreak{}a\allowbreak{}m\allowbreak{}e\allowbreak{}>\allowbreak{}.\allowbreak{}p\allowbreak{}n\allowbreak{}g}
\end{itemize}
\textbf{Look at each screen's PNG before you implement that screen.} Open the image itself --- listing the file, reading its bytes, or working from the JSON and the description alone is not the same thing. The PNG is what the screen has to look like; the JSON and the description tell you what is on it, not how it reads.

Per-screen inputs: \textbf{{\promptfont <\allowbreak{}s\allowbreak{}c\allowbreak{}r\allowbreak{}e\allowbreak{}e\allowbreak{}n\allowbreak{}>\allowbreak{}.\allowbreak{}p\allowbreak{}n\allowbreak{}g}} (rendered mockup, visual ground truth), \textbf{{\promptfont <\allowbreak{}s\allowbreak{}c\allowbreak{}r\allowbreak{}e\allowbreak{}e\allowbreak{}n\allowbreak{}>\allowbreak{}.\allowbreak{}j\allowbreak{}s\allowbreak{}o\allowbreak{}n}} (Figma structure + bbox), the per-screen bullet in {\promptfont d\allowbreak{}e\allowbreak{}s\allowbreak{}c\allowbreak{}r\allowbreak{}i\allowbreak{}p\allowbreak{}t\allowbreak{}i\allowbreak{}o\allowbreak{}n\allowbreak{}.\allowbreak{}m\allowbreak{}d} (intent + inline {\promptfont <\allowbreak{}s\allowbreak{}n\allowbreak{}a\allowbreak{}k\allowbreak{}e\allowbreak{}\_\allowbreak{}c\allowbreak{}a\allowbreak{}s\allowbreak{}e\allowbreak{}>} anchor markers), and \textbf{{\promptfont .\allowbreak{}.\allowbreak{}/\allowbreak{}i\allowbreak{}n\allowbreak{}p\allowbreak{}u\allowbreak{}t\allowbreak{}s\allowbreak{}/\allowbreak{}d\allowbreak{}e\allowbreak{}s\allowbreak{}i\allowbreak{}g\allowbreak{}n\allowbreak{}\_\allowbreak{}s\allowbreak{}y\allowbreak{}s\allowbreak{}t\allowbreak{}e\allowbreak{}m\allowbreak{}/\allowbreak{}u\allowbreak{}n\allowbreak{}i\allowbreak{}t\allowbreak{}s\allowbreak{}/\allowbreak{}<\allowbreak{}C\allowbreak{}a\allowbreak{}t\allowbreak{}e\allowbreak{}g\allowbreak{}o\allowbreak{}r\allowbreak{}y\allowbreak{}>\allowbreak{}/\allowbreak{}<\allowbreak{}n\allowbreak{}a\allowbreak{}m\allowbreak{}e\allowbreak{}>\allowbreak{}.\allowbreak{}p\allowbreak{}n\allowbreak{}g}} (each button / icon / input rendered on its own --- the source of truth for component styling; build these as real widgets).

\PromptSubheading{Coordinates in {\promptfont <\allowbreak{}s\allowbreak{}c\allowbreak{}r\allowbreak{}e\allowbreak{}e\allowbreak{}n\allowbreak{}>\allowbreak{}.\allowbreak{}j\allowbreak{}s\allowbreak{}o\allowbreak{}n} --- use {\promptfont f\allowbreak{}r\allowbreak{}a\allowbreak{}m\allowbreak{}e\allowbreak{}\_\allowbreak{}b\allowbreak{}b\allowbreak{}o\allowbreak{}x}, not {\promptfont a\allowbreak{}b\allowbreak{}s\allowbreak{}o\allowbreak{}l\allowbreak{}u\allowbreak{}t\allowbreak{}e\allowbreak{}B\allowbreak{}o\allowbreak{}u\allowbreak{}n\allowbreak{}d\allowbreak{}i\allowbreak{}n\allowbreak{}g\allowbreak{}B\allowbreak{}o\allowbreak{}x}}
Every node carries three boxes. Only one of them is a layout coordinate:

\begin{itemize}
\item \textbf{{\promptfont f\allowbreak{}r\allowbreak{}a\allowbreak{}m\allowbreak{}e\allowbreak{}\_\allowbreak{}b\allowbreak{}b\allowbreak{}o\allowbreak{}x}} --- {\promptfont \{\allowbreak{}x\allowbreak{},\allowbreak{} \allowbreak{}y\allowbreak{},\allowbreak{} \allowbreak{}w\allowbreak{},\allowbreak{} \allowbreak{}h\allowbreak{}\}} in \textbf{dp, origin = the screen's top-left}.
This is the one to build from.

\item {\promptfont b\allowbreak{}b\allowbreak{}o\allowbreak{}x\allowbreak{}\_\allowbreak{}p\allowbreak{}n\allowbreak{}g} --- the same box in PNG pixels ({\promptfont f\allowbreak{}r\allowbreak{}a\allowbreak{}m\allowbreak{}e\allowbreak{}\_\allowbreak{}b\allowbreak{}b\allowbreak{}o\allowbreak{}x} x {\promptfont p\allowbreak{}n\allowbreak{}g\allowbreak{}\_\allowbreak{}s\allowbreak{}c\allowbreak{}a\allowbreak{}l\allowbreak{}e}, always 2
here). Use it to line a node up against the mockup image.

\item {\promptfont a\allowbreak{}b\allowbreak{}s\allowbreak{}o\allowbreak{}l\allowbreak{}u\allowbreak{}t\allowbreak{}e\allowbreak{}B\allowbreak{}o\allowbreak{}u\allowbreak{}n\allowbreak{}d\allowbreak{}i\allowbreak{}n\allowbreak{}g\allowbreak{}B\allowbreak{}o\allowbreak{}x} --- Figma \textbf{canvas} space. Its origin is wherever the
designer parked the frame on the canvas, so it is a large arbitrary number (e.g. {\promptfont x\allowbreak{}:\allowbreak{} \allowbreak{}-\allowbreak{}4\allowbreak{}5\allowbreak{}0\allowbreak{}1\allowbreak{},\allowbreak{} \allowbreak{}y\allowbreak{}:\allowbreak{} \allowbreak{}-\allowbreak{}1\allowbreak{}4\allowbreak{}1\allowbreak{}5\allowbreak{}3}). \textbf{Never place a widget at these coordinates} --- the UI lands off-screen.

\end{itemize}
The file's top-level {\promptfont f\allowbreak{}i\allowbreak{}g\allowbreak{}m\allowbreak{}a\allowbreak{}\_\allowbreak{}m\allowbreak{}e\allowbreak{}t\allowbreak{}a} states {\promptfont o\allowbreak{}r\allowbreak{}i\allowbreak{}g\allowbreak{}i\allowbreak{}n\allowbreak{}\_\allowbreak{}x\allowbreak{} \allowbreak{}/\allowbreak{} \allowbreak{}o\allowbreak{}r\allowbreak{}i\allowbreak{}g\allowbreak{}i\allowbreak{}n\allowbreak{}\_\allowbreak{}y\allowbreak{} \allowbreak{}/\allowbreak{} \allowbreak{}f\allowbreak{}i\allowbreak{}g\allowbreak{}m\allowbreak{}a\allowbreak{}\_\allowbreak{}w\allowbreak{} \allowbreak{}/\allowbreak{} \allowbreak{}f\allowbreak{}i\allowbreak{}g\allowbreak{}m\allowbreak{}a\allowbreak{}\_\allowbreak{}h\allowbreak{} \allowbreak{}/\allowbreak{} \allowbreak{}p\allowbreak{}n\allowbreak{}g\allowbreak{}\_\allowbreak{}s\allowbreak{}c\allowbreak{}a\allowbreak{}l\allowbreak{}e}, so {\promptfont f\allowbreak{}r\allowbreak{}a\allowbreak{}m\allowbreak{}e\allowbreak{}\_\allowbreak{}b\allowbreak{}b\allowbreak{}o\allowbreak{}x\allowbreak{} \allowbreak{}=\allowbreak{} \allowbreak{}a\allowbreak{}b\allowbreak{}s\allowbreak{}o\allowbreak{}l\allowbreak{}u\allowbreak{}t\allowbreak{}e\allowbreak{}B\allowbreak{}o\allowbreak{}u\allowbreak{}n\allowbreak{}d\allowbreak{}i\allowbreak{}n\allowbreak{}g\allowbreak{}B\allowbreak{}o\allowbreak{}x\allowbreak{} \allowbreak{}-\allowbreak{} \allowbreak{}(\allowbreak{}o\allowbreak{}r\allowbreak{}i\allowbreak{}g\allowbreak{}i\allowbreak{}n\allowbreak{}\_\allowbreak{}x\allowbreak{},\allowbreak{} \allowbreak{}o\allowbreak{}r\allowbreak{}i\allowbreak{}g\allowbreak{}i\allowbreak{}n\allowbreak{}\_\allowbreak{}y\allowbreak{})} if you ever need to re-derive it. {\promptfont f\allowbreak{}i\allowbreak{}g\allowbreak{}m\allowbreak{}a\allowbreak{}\_\allowbreak{}w\allowbreak{} \allowbreak{}x\allowbreak{} \allowbreak{}f\allowbreak{}i\allowbreak{}g\allowbreak{}m\allowbreak{}a\allowbreak{}\_\allowbreak{}h} is the screen's design size in dp and varies per task (360, 375, 393, 430 wide); build to that width, do not assume 360.

\textbf{Read them before writing each screen.} Reproduce faithfully --- the PNG is the reference for how the screen must look. One component per named element; match hierarchy; coordinates are Figma \textbf{dp} ({\promptfont f\allowbreak{}i\allowbreak{}g\allowbreak{}m\allowbreak{}a\allowbreak{}\_\allowbreak{}m\allowbreak{}e\allowbreak{}t\allowbreak{}a} gives {\promptfont f\allowbreak{}i\allowbreak{}g\allowbreak{}m\allowbreak{}a\allowbreak{}\_\allowbreak{}w\allowbreak{} \allowbreak{}\ensuremath{\times}\allowbreak{} \allowbreak{}f\allowbreak{}i\allowbreak{}g\allowbreak{}m\allowbreak{}a\allowbreak{}\_\allowbreak{}h}, build in dp); text from the JSON's {\promptfont c\allowbreak{}h\allowbreak{}a\allowbreak{}r\allowbreak{}a\allowbreak{}c\allowbreak{}t\allowbreak{}e\allowbreak{}r\allowbreak{}s}; no silent compression. The \emph{values} (names, prices, items) come from the backend; the \emph{layout/styling} comes from the mockup + design system.

\PromptHeading{ELEMENT IDENTIFIER CONTRACT}
{\promptfont d\allowbreak{}e\allowbreak{}s\allowbreak{}c\allowbreak{}r\allowbreak{}i\allowbreak{}p\allowbreak{}t\allowbreak{}i\allowbreak{}o\allowbreak{}n\allowbreak{}.\allowbreak{}m\allowbreak{}d} marks anchors as {\promptfont <\allowbreak{}s\allowbreak{}n\allowbreak{}a\allowbreak{}k\allowbreak{}e\allowbreak{}\_\allowbreak{}c\allowbreak{}a\allowbreak{}s\allowbreak{}e\allowbreak{}>} next to their element and lists them per screen under {\promptfont -\allowbreak{} \allowbreak{}*\allowbreak{}*\allowbreak{}A\allowbreak{}n\allowbreak{}c\allowbreak{}h\allowbreak{}o\allowbreak{}r\allowbreak{}s\allowbreak{}:\allowbreak{}*\allowbreak{}*}. Give each such element a stable identifier equal to the exact marker string --- the ordinary way an accessible Android app labels controls. It belongs on the \textbf{real, visible widget} it names, placed where the mockup puts it.

\begin{enumerate}
\item \textbf{Exact string}, no brackets, no prefixes/suffixes/camelCase: {\promptfont <\allowbreak{}f\allowbreak{}i\allowbreak{}l\allowbreak{}t\allowbreak{}e\allowbreak{}r\allowbreak{}\_\allowbreak{}s\allowbreak{}e\allowbreak{}a\allowbreak{}r\allowbreak{}c\allowbreak{}h\allowbreak{}>} \ensuremath{\rightarrow} {\promptfont f\allowbreak{}i\allowbreak{}l\allowbreak{}t\allowbreak{}e\allowbreak{}r\allowbreak{}\_\allowbreak{}s\allowbreak{}e\allowbreak{}a\allowbreak{}r\allowbreak{}c\allowbreak{}h}.
\item Per stack: Views {\promptfont a\allowbreak{}n\allowbreak{}d\allowbreak{}r\allowbreak{}o\allowbreak{}i\allowbreak{}d\allowbreak{}:\allowbreak{}i\allowbreak{}d\allowbreak{}=\allowbreak{}"\allowbreak{}@\allowbreak{}+\allowbreak{}i\allowbreak{}d\allowbreak{}/\allowbreak{}f\allowbreak{}i\allowbreak{}l\allowbreak{}t\allowbreak{}e\allowbreak{}r\allowbreak{}\_\allowbreak{}s\allowbreak{}e\allowbreak{}a\allowbreak{}r\allowbreak{}c\allowbreak{}h\allowbreak{}"}; Compose {\promptfont M\allowbreak{}o\allowbreak{}d\allowbreak{}i\allowbreak{}f\allowbreak{}i\allowbreak{}e\allowbreak{}r\allowbreak{}.\allowbreak{}t\allowbreak{}e\allowbreak{}s\allowbreak{}t\allowbreak{}T\allowbreak{}a\allowbreak{}g\allowbreak{}(\allowbreak{}"\allowbreak{}f\allowbreak{}i\allowbreak{}l\allowbreak{}t\allowbreak{}e\allowbreak{}r\allowbreak{}\_\allowbreak{}s\allowbreak{}e\allowbreak{}a\allowbreak{}r\allowbreak{}c\allowbreak{}h\allowbreak{}"\allowbreak{})} \textbf{plus} {\promptfont M\allowbreak{}o\allowbreak{}d\allowbreak{}i\allowbreak{}f\allowbreak{}i\allowbreak{}e\allowbreak{}r\allowbreak{}.\allowbreak{}s\allowbreak{}e\allowbreak{}m\allowbreak{}a\allowbreak{}n\allowbreak{}t\allowbreak{}i\allowbreak{}c\allowbreak{}s\allowbreak{} \allowbreak{}\{\allowbreak{} \allowbreak{}t\allowbreak{}e\allowbreak{}s\allowbreak{}t\allowbreak{}T\allowbreak{}a\allowbreak{}g\allowbreak{}s\allowbreak{}A\allowbreak{}s\allowbreak{}R\allowbreak{}e\allowbreak{}s\allowbreak{}o\allowbreak{}u\allowbreak{}r\allowbreak{}c\allowbreak{}e\allowbreak{}I\allowbreak{}d\allowbreak{} \allowbreak{}=\allowbreak{} \allowbreak{}t\allowbreak{}r\allowbreak{}u\allowbreak{}e\allowbreak{} \allowbreak{}\}} at the screen root; RN {\promptfont t\allowbreak{}e\allowbreak{}s\allowbreak{}t\allowbreak{}I\allowbreak{}D}; Flutter {\promptfont S\allowbreak{}e\allowbreak{}m\allowbreak{}a\allowbreak{}n\allowbreak{}t\allowbreak{}i\allowbreak{}c\allowbreak{}s\allowbreak{}(\allowbreak{}i\allowbreak{}d\allowbreak{}e\allowbreak{}n\allowbreak{}t\allowbreak{}i\allowbreak{}f\allowbreak{}i\allowbreak{}e\allowbreak{}r\allowbreak{}:\allowbreak{})}.
\item Reuse the same anchor across screens for the same logical element (e.g. the bottom tab bar). Repeated list items each carry the same id --- position tells them apart; don't append indices.
\item \textbf{Only the marked anchors, on their real elements} --- a handful per screen. Never attach an anchor id to a hidden, empty, or zero-size placeholder to make it "present".
\end{enumerate}
\PromptHeading{VERIFICATION}
Before declaring done, verify ALL:

\begin{enumerate}
\item \textbf{Backend up}: {\promptfont d\allowbreak{}o\allowbreak{}c\allowbreak{}k\allowbreak{}e\allowbreak{}r\allowbreak{} \allowbreak{}c\allowbreak{}o\allowbreak{}m\allowbreak{}p\allowbreak{}o\allowbreak{}s\allowbreak{}e\allowbreak{} \allowbreak{}u\allowbreak{}p\allowbreak{} \allowbreak{}-\allowbreak{}-\allowbreak{}b\allowbreak{}u\allowbreak{}i\allowbreak{}l\allowbreak{}d\allowbreak{} \allowbreak{}-\allowbreak{}d} \ensuremath{\rightarrow} backend healthy on {\promptfont :\allowbreak{}8\allowbreak{}0\allowbreak{}8\allowbreak{}0}; hitting a couple of endpoints ({\promptfont c\allowbreak{}u\allowbreak{}r\allowbreak{}l\allowbreak{} \allowbreak{}l\allowbreak{}o\allowbreak{}c\allowbreak{}a\allowbreak{}l\allowbreak{}h\allowbreak{}o\allowbreak{}s\allowbreak{}t\allowbreak{}:\allowbreak{}8\allowbreak{}0\allowbreak{}8\allowbreak{}0\allowbreak{}/\allowbreak{}.\allowbreak{}.\allowbreak{}.}) returns seeded JSON.
\item \textbf{Clean app build}: {\promptfont r\allowbreak{}m\allowbreak{} \allowbreak{}-\allowbreak{}r\allowbreak{}f\allowbreak{} \allowbreak{}o\allowbreak{}u\allowbreak{}t\allowbreak{} \allowbreak{}\&\allowbreak{}\&\allowbreak{} \allowbreak{}.\allowbreak{}/\allowbreak{}b\allowbreak{}u\allowbreak{}i\allowbreak{}l\allowbreak{}d\allowbreak{}.\allowbreak{}s\allowbreak{}h} \ensuremath{\rightarrow} exit 0, {\promptfont o\allowbreak{}u\allowbreak{}t\allowbreak{}/\allowbreak{}a\allowbreak{}p\allowbreak{}p\allowbreak{}-\allowbreak{}d\allowbreak{}e\allowbreak{}b\allowbreak{}u\allowbreak{}g\allowbreak{}.\allowbreak{}a\allowbreak{}p\allowbreak{}k} exists.
\item \textbf{Cold install + launch} with the backend running: {\promptfont a\allowbreak{}d\allowbreak{}b\allowbreak{} \allowbreak{}i\allowbreak{}n\allowbreak{}s\allowbreak{}t\allowbreak{}a\allowbreak{}l\allowbreak{}l\allowbreak{} \allowbreak{}-\allowbreak{}r\allowbreak{} \allowbreak{}o\allowbreak{}u\allowbreak{}t\allowbreak{}/\allowbreak{}a\allowbreak{}p\allowbreak{}p\allowbreak{}-\allowbreak{}d\allowbreak{}e\allowbreak{}b\allowbreak{}u\allowbreak{}g\allowbreak{}.\allowbreak{}a\allowbreak{}p\allowbreak{}k}, then launch renders the launch screen, and screens load \textbf{real backend data} (not empty, no permanent spinner).
\item \textbf{Every deep link works}: for each {\promptfont \#\allowbreak{}\#\allowbreak{} \allowbreak{}S\allowbreak{}c\allowbreak{}r\allowbreak{}e\allowbreak{}e\allowbreak{}n\allowbreak{}s} row, {\promptfont a\allowbreak{}d\allowbreak{}b\allowbreak{} \allowbreak{}s\allowbreak{}h\allowbreak{}e\allowbreak{}l\allowbreak{}l\allowbreak{} \allowbreak{}a\allowbreak{}m\allowbreak{} \allowbreak{}s\allowbreak{}t\allowbreak{}a\allowbreak{}r\allowbreak{}t\allowbreak{} \allowbreak{}-\allowbreak{}W\allowbreak{} \allowbreak{}-\allowbreak{}a\allowbreak{} \allowbreak{}a\allowbreak{}n\allowbreak{}d\allowbreak{}r\allowbreak{}o\allowbreak{}i\allowbreak{}d\allowbreak{}.\allowbreak{}i\allowbreak{}n\allowbreak{}t\allowbreak{}e\allowbreak{}n\allowbreak{}t\allowbreak{}.\allowbreak{}a\allowbreak{}c\allowbreak{}t\allowbreak{}i\allowbreak{}o\allowbreak{}n\allowbreak{}.\allowbreak{}V\allowbreak{}I\allowbreak{}E\allowbreak{}W\allowbreak{} \allowbreak{}-\allowbreak{}d\allowbreak{} \allowbreak{}"\allowbreak{}<\allowbreak{}l\allowbreak{}i\allowbreak{}n\allowbreak{}k\allowbreak{}>\allowbreak{}"\allowbreak{} \allowbreak{}<\allowbreak{}p\allowbreak{}k\allowbreak{}g\allowbreak{}>} returns {\promptfont S\allowbreak{}t\allowbreak{}a\allowbreak{}t\allowbreak{}u\allowbreak{}s\allowbreak{}:\allowbreak{} \allowbreak{}o\allowbreak{}k} and shows the right screen, populated.
\item \textbf{Visual audit per screen}: every named element / Figma node / mockup section has an implementation; styling matches the design system.
\item \textbf{Identifiers present and exact} --- deep-link to a screen, {\promptfont u\allowbreak{}i\allowbreak{}a\allowbreak{}u\allowbreak{}t\allowbreak{}o\allowbreak{}m\allowbreak{}a\allowbreak{}t\allowbreak{}o\allowbreak{}r\allowbreak{} \allowbreak{}d\allowbreak{}u\allowbreak{}m\allowbreak{}p}, diff resource-ids against the markers in {\promptfont d\allowbreak{}e\allowbreak{}s\allowbreak{}c\allowbreak{}r\allowbreak{}i\allowbreak{}p\allowbreak{}t\allowbreak{}i\allowbreak{}o\allowbreak{}n\allowbreak{}.\allowbreak{}m\allowbreak{}d}; fix any missing/misspelled.
\item \textbf{In-app navigation complete} --- from the launch screen every {\promptfont s\allowbreak{}i\allowbreak{}t\allowbreak{}e\allowbreak{}m\allowbreak{}a\allowbreak{}p\allowbreak{}.\allowbreak{}m\allowbreak{}d} screen is reachable by tapping.
\end{enumerate}
Fix any failure. Do not declare done until 1--7 all pass.

\PromptHeading{FAILURE HANDLING}
If a requirement is ambiguous, implement the simplest reasonable interpretation and leave a code comment. If a screen's design is unreadable from the PNG, implement your best guess from the Figma JSON --- never skip a screen.

\emph{--- End of System Prompt. Task description appended below. ---}

\end{SystemPromptBox}

\clearpage
\section{Model-level Workflow Profiles}
\label{app:workflow-trajectories}
The complete profiles disaggregate the harness overview in
Figure~\ref{fig:workflow-trajectories}. Each rhythm profile contains ten progress
bins and seven action categories. Each transition matrix is normalized by its
source action over the same seven categories, including VERIFY.

\begin{figure}[p]
\centering
\includegraphics[width=\linewidth,height=0.84\textheight,keepaspectratio]{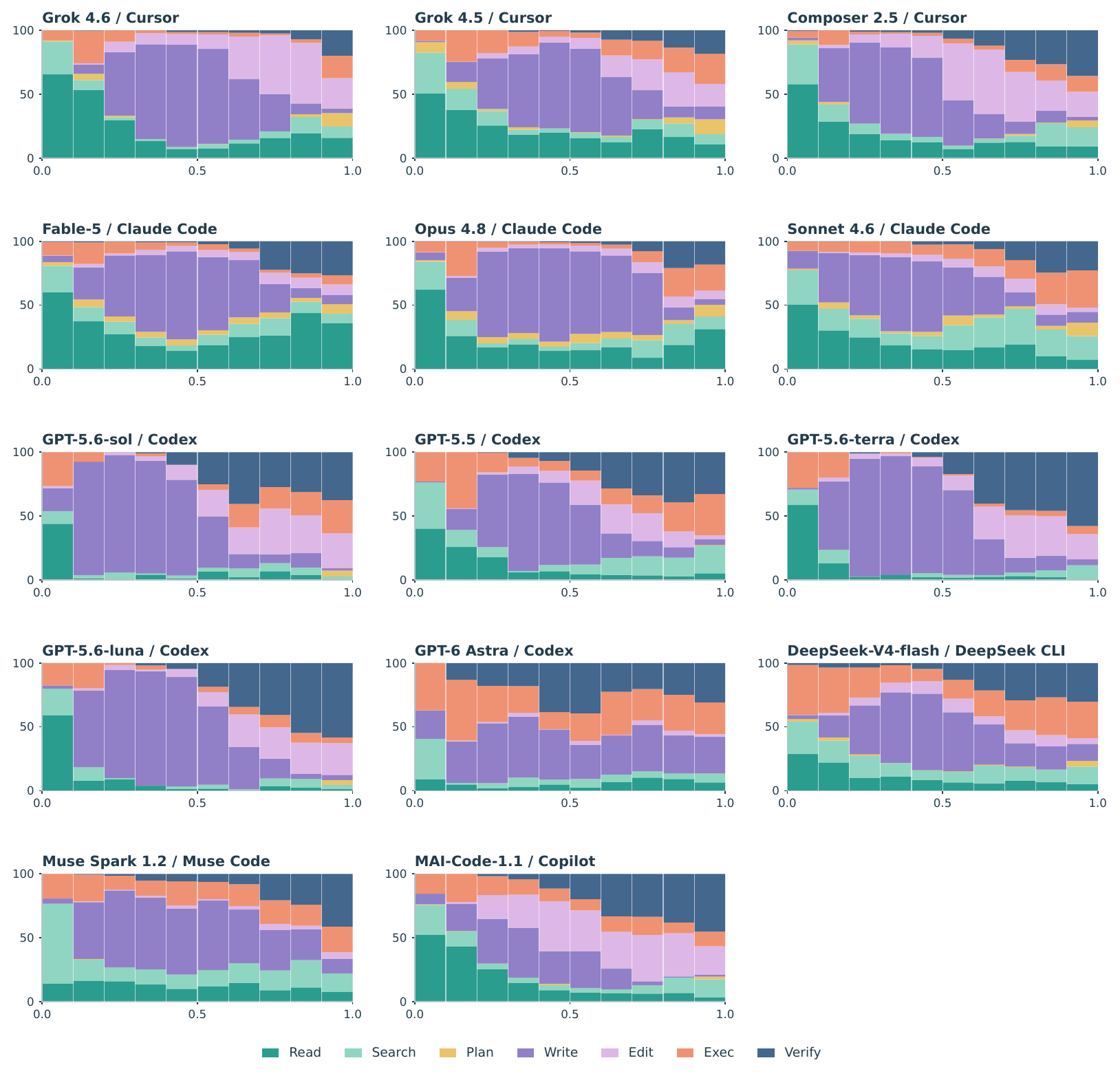}
\caption{Workflow rhythm for all 14 models. Colors match the harness overview;
vertical axes show action shares in percent and horizontal axes show normalized
run progress. Shares are normalized within each model/bin over the recorded categories;
model profiles are then averaged equally within each harness in the main figure.}
\label{fig:workflow-model-rhythm}
\end{figure}

\begin{figure}[p]
\centering
\includegraphics[width=\linewidth,height=0.84\textheight,keepaspectratio]{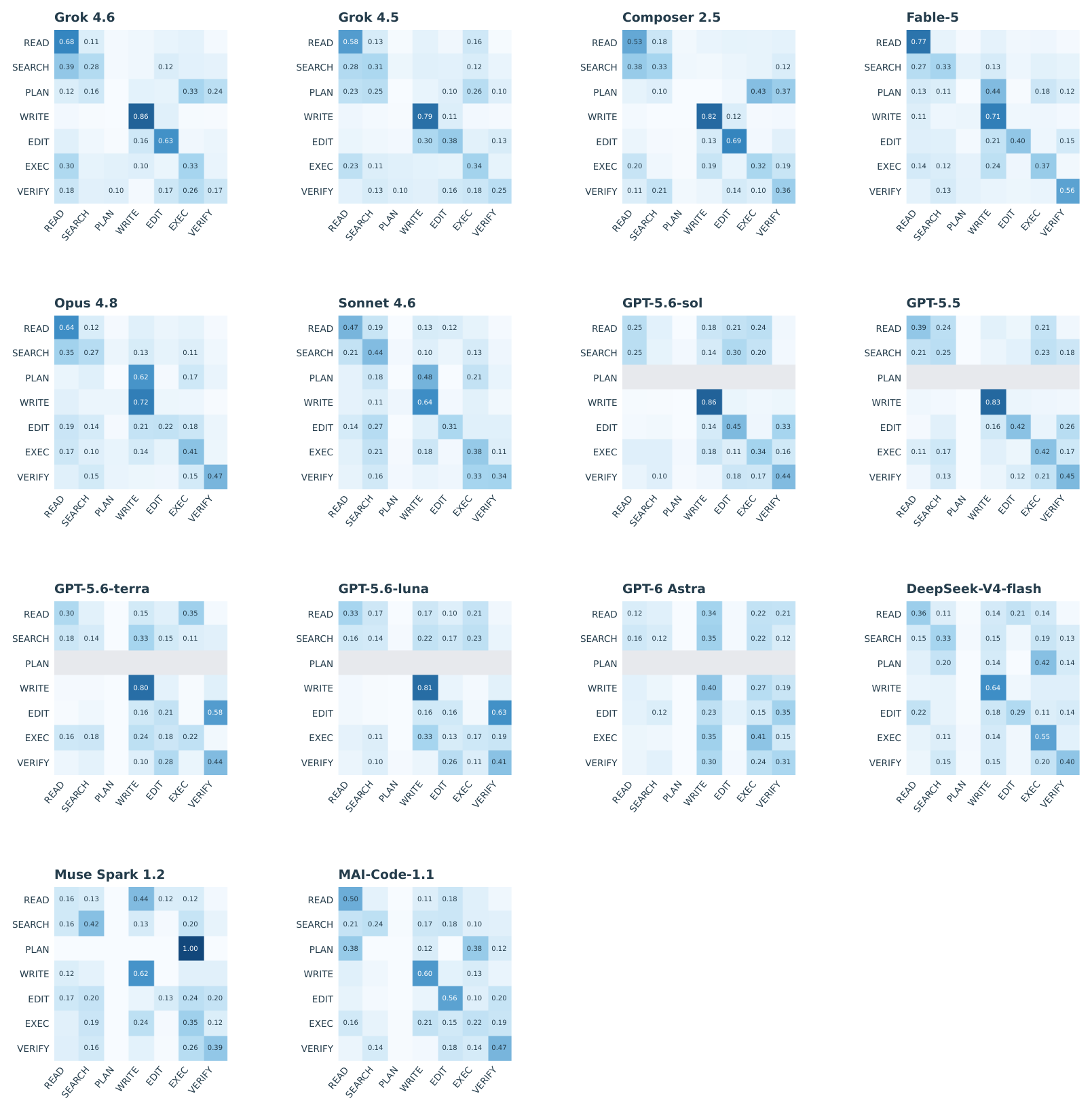}
\caption{Action-transition matrices for all 14 models. Rows indicate source
categories and columns indicate destination categories. Values below 0.10 are
unannotated; gray rows indicate no observed source transitions.}
\label{fig:workflow-model-transitions}
\end{figure}

\clearpage
\section{Input and Stack Contrasts}
\label{app:condition-contrasts}

The condition study covers GPT-5.5 and Sonnet 4.6. We estimate contrasts from
agent--task pairs after averaging repeated runs within each condition.
Percentile intervals use 10,000 bootstrap resamples \citep{efron1986bootstrap};
two-sided tests use 10,000 paired sign flips \citep{hemerik2018exact}.
Tool-use and token values in Section~\ref{sec:input-stack-analysis} are
equal-weight averages of per-agent condition medians; token workload counts
non-cached input and output tokens.

Under a free stack, adding screenshots yields $\Delta S=+0.0071$
($[-0.0920,+0.1073]$), and adding structure JSON yields $\Delta S=-0.0228$
($[-0.1053,+0.0614]$). Relaxing the stack constraint yields $\Delta S=+0.0006$
($[-0.0737,+0.0751]$) at the screenshot tier and $+0.0200$
($[-0.0404,+0.0877]$) at the structure tier. At the screenshot tier, free stack
choice relative to the fixed stack yields $\Delta L=-0.0888$
($[-0.2203,+0.0282]$) and $\Delta B=-0.0122$ ($[-0.1004,+0.0734]$).
Comparisons involving $C_5$ were collected in a different batch with a
different harness build and are therefore not included in these contrasts.

\clearpage
\section{Scores Across Harness Versions}
\label{app:harness-versions}

\begin{figure}[htbp]
\centering
\includegraphics[width=\linewidth]{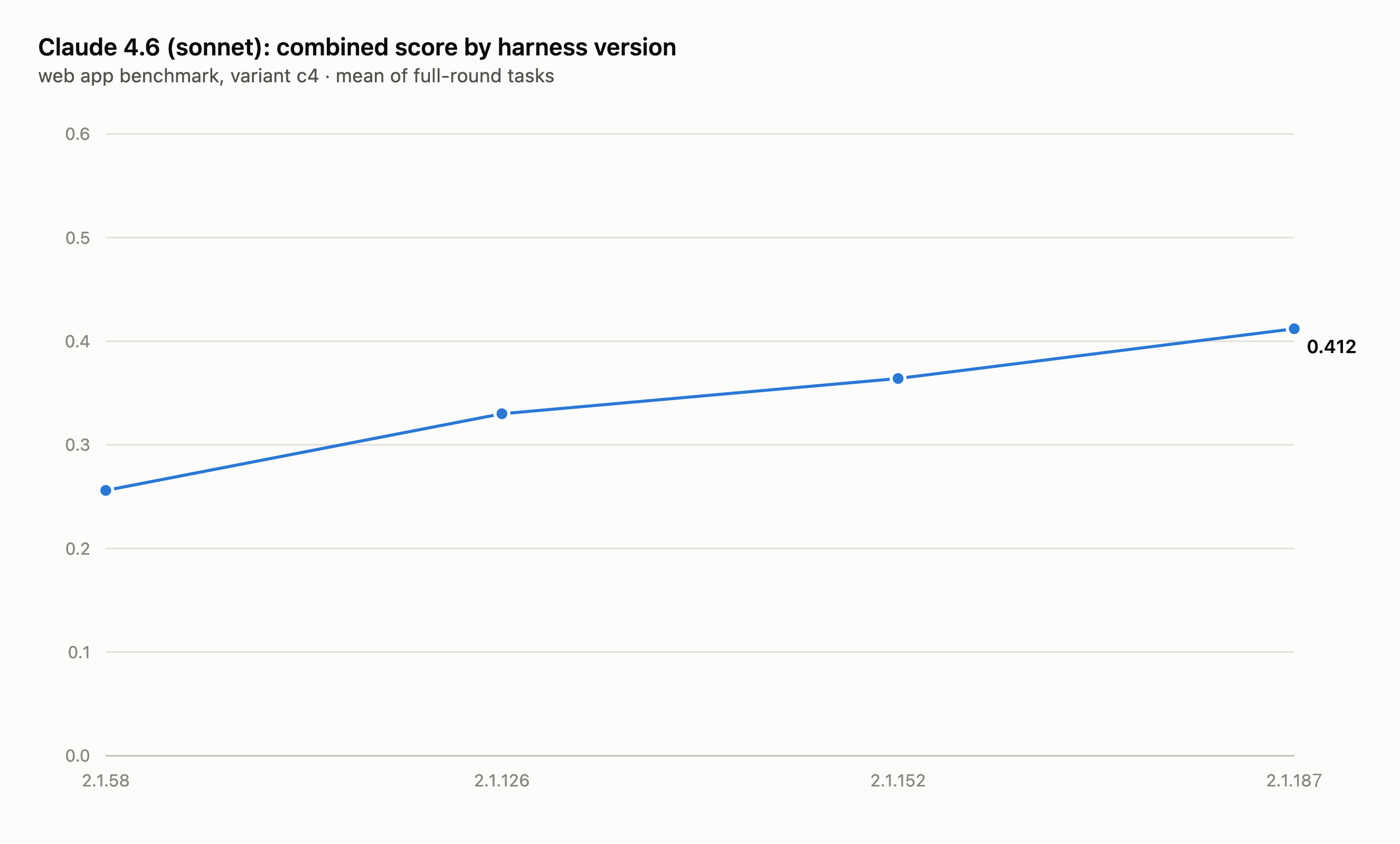}
\caption{Sonnet 4.6 joint score across harness versions 2.1.58, 2.1.126,
2.1.152, and 2.1.187 on the Web benchmark under $C_4$. Each point reports
the mean over full-round tasks.}
\label{fig:sonnet-harness-score}
\end{figure}

Figure~\ref{fig:sonnet-harness-score} compares Sonnet 4.6 across four harness
versions on the Web benchmark under $C_4$. The mean joint score increases across
the versions shown, reaching $S=0.412$ with version 2.1.187, the version used in
Table~\ref{tab:evaluation-web-mobile}.

\end{document}